\documentclass{mn2e}
\usepackage{psfig,graphicx}

\begin{document}

\title
[Nonlinear pulsations in differentially rotating neutron stars]
{Nonlinear pulsations in differentially rotating neutron stars:
Mass-shedding-induced damping and splitting of the fundamental mode}

\author[Stergioulas, Apostolatos and Font]
{Nikolaos Stergioulas$^{(1)}$,
Theocharis A.~Apostolatos$^{(2)}$,
and Jos\'e A.~Font$^{(3)}$ \\
$^(1)$Department of Physics, Aristotle University of Thessaloniki,
Thessaloniki 54124, Greece \\
$^(2)$Department of Physics, National and Kapodistrian University
of Athens, Panepistimiopolis, Athens 15783, Greece \\
$^(3)$Departamento de Astronom\'{\i}a y Astrof\'{\i}sica, Universidad
de Valencia, Dr. Moliner 50, 46100 Burjassot (Valencia), Spain
}

\maketitle

\begin{abstract}
  We study small-amplitude, nonlinear pulsations of uniformly and
  differentially rotating neutron stars employing a two-dimensional
  evolution code for general-relativistic hydrodynamics. Using Fourier
  transforms at several points inside the star, both the
  eigenfrequencies and two-dimensional eigenfunctions of pulsations
  are extracted. The centrifugal forces and the degree of differential
  rotation have significant effects on the mode-eigenfunction. We find
  that near the mass-shedding limit, the pulsations are damped due to
  shocks forming at the surface of the star. This new damping
  mechanism may set a small saturation amplitude for modes that are
  unstable to the emission of gravitational waves.  After correcting
  for the assumption of the Cowling approximation (used in our
  numerical code), we construct empirical relations that predict the
  range of gravitational-wave frequencies from quasi-periodic
  post-bounce oscillations in the core collapse of massive stars.  We also
  find that the fundamental quasi-radial mode is split, at least in
  the Cowling approximation and mainly in differentially rotating
  stars, into two different sequences.
\end{abstract}

\section{Introduction}

When neutron stars are formed as a result of a violent event (a core
collapse, an accretion-induced collapse, or a binary neutron star
merger), they are expected to pulsate nonlinearly before various
dissipative effects, such as viscosity, magnetic braking and shock
formation, settle them down to a near-equilibrium configuration.  All
these violent events could lead to high rotational rates for the final
objects, due to angular momentum conservation.  Moreover, the final
object is expected to rotate differentially, since the initial
distribution of angular momentum has to be redistributed after
collapse or merger. Several axisymmetric and nonaxisymmetric pulsation
modes are expected to be excited during neutron star formation and
could be important sources of gravitational waves.  In addition, some modes
could become unstable to the emission of gravitational radiation. The
successful detection and identification of pulsation modes requires a
detailed understanding of the eigenfrequency and eigenfunction of each
mode-sequence, especially since pulsation modes are high-frequency
gravitational-wave sources. In addition, it is necessary to have a
good understanding of the various damping mechanisms that
will operate.

There are several fundamental questions that have to be resolved
before we can realistically hope for detection of gravitational waves
from neutron star pulsations: {\it Can neutron stars attain high
  angular momentum at birth? How strongly differential is the initial
  rotational profile? What kind of modes are excited?  What is the
  initial amplitude for each mode? What is the frequency for each
  mode?  Which mechanisms dampen the pulsations and on what
  timescale?} To date, none of the above questions has a definitive
answer, but some partial understanding is emerging.

{\it Initial rotation rate:} Naively, the collapse of a rotating
stellar core should lead to an extremely rapidly rotating neutron
star, just by the argument of angular momentum conservation. This
picture is, however, made more complicated due to the presence of
magnetic fields, wind-mass loss, centrifugal hangup during collapse,
fall-back accretion and pre- or post-collapse binary interactions. In
recent work by Heger et al.~(2003) who consider evolutionary sequences
of realistic rotating pre-collapse cores, it is suggested that the
initial rotational period of a neutron star can indeed be of the order
of 1ms or less, if the magnetic field is neglected.  However, if the
star passes through a red-supergiant phase, then the dynamo model for
magnetic braking of the angular velocity (Spruit, 2002), increases the
initial period of proto-neutron stars to several milliseconds.
Wind-mass loss during a Wolf-Rayet/helium-star phase (Heger \&
Woosley, 2003) is another mechanism for slowing down the core of an
evolved star.  Nevertheless, more rapidly rotating proto-neutron stars
could be obtained through other formation scenarios, such as
fall-back accretion (see Watts \& Andersson, 2002, and references
therein), pre-supernova binary interactions (Pfahl et al.  2002;
Ivanova \& Podsiadlowski, 2003; Podsiadlowski et al. 2003) and
post-supernova accretion from a binary companion (Langer et al. 2003).
A new scenario for millisecond-pulsar birth in globular clusters is
the merger of binary white dwarfs (Middleditch, 2003).

Recently, Villain et al. (2003) have studied in detail the
quasi-stationary evolution from core collapse to the formation of a
neutron star and find that in most cases an initially hot
proto-neutron star contracts and spins-up on a time-scale of seconds,
until reaching high rotation rates with $T/|W|>0.1-0.2$, (where $T$ is
the kinetic energy and $W$ is the gravitational binding energy of the
star). If the initial core is rapidly rotating, centrifugal hangup can
occur during collapse. The rotational evolution of a proto-neutron
star can also be affected by the details of the high-density equation
of state (EOS hereafter), e.g., the appearance of hyperons in the core
can yield a very rapidly rotating, but metastable, compact star (Yuan
\& Heyl, 2003).  Axisymmetric core collapse simulations in general
relativity (Dimmelmeier, Font \& M\"{u}ller, 2001, 2002a, 2002b; Shibata,
2003) and in Newtonian gravity but with realistic initial conditions
(Ott et al.  2003; Kotake, Tamada \& Sato, 2003; see also Fryer et
al. 2002 and references therein) have also shown the formation of
rapidly rotating neutron stars.


{\it Differential rotation:} The core collapse studies by Dimmelmeier
et al. (2001, 2002a, 2002b), and a quasi-equilibrium treatment by Liu
\& Lindblom (2000) in the case of accretion-induced collapse of white
dwarfs, have shown that rapidly rotating proto-neutron stars are born
with a modest degree of differential rotation. A recent detailed
analysis by Villain et al. (2003) shows that the length-scale on which
the angular velocity changes within the star is of the order of 7-10
km.  Hypermassive neutron stars created in a binary merger also have
moderate differential rotation, with a somewhat shorter length-scale
than in core collapse (see e.g., Shibata \& Ury\={u}, 2000).  A
fundamental question about the rotational profile is how quickly it
will become uniform. Shapiro (2000) and Cook, Shapiro \& Stephens
(2003) have suggested that magnetic braking could drive the star to
uniform rotation on a timescale of only seconds.  A more recent study
by Liu \& Shapiro (2003) shows that a magnetic field will cause
turbulent motions in a differentially rotating star, driving it to
uniform rotation.  It has also been suggested that a differentially
rotating, viscous proto-neutron star will be brought to uniform
rotation due to turbulent mixing on a much shorter timescale (Hegyi,
1977), however, more detailed computations, including realistic
composition gradients, are required in order to obtain better
estimates.

{\it Mode excitation:} In the rotational core collapse simulations by
Dimmelmeier et al. (2001, 2002a, 2002b) the proto-neutron star quickly
settles into quasi-equilibrium after core bounce.  Still, pulsations
are excited and survive for several oscillation periods. The dominant
pulsation modes that are expected to be excited in rotational core
collapse are the quasi-radial ($l=m=0$) mode and the quadrupole ($l=2,
m=0$) mode.  Due to rotational couplings of nonradial terms in its
eigenfunction, the quasi-radial mode becomes a strong emitter of
gravitational waves in rapidly rotating stars (it could, in fact,
become the dominant mode in which gravitational waves are emitted).
The amplitude of the oscillations in the density is estimated
as several percent at the center of the star.  Newtonian simulations
(e.g., M\"onchmeyer et al. 1991; Zwerger \& M\"uller, 1997) also
reached similar conclusions (for a recent review of core collapse
simulations see New, 2002).  In the simulations of binary neutron
star mergers by Shibata and Ury\={u} (2000, 2002), quasi-periodic
oscillations are excited, through which strong gravitational waves are
emitted. The frequency of these oscillations suggest that they could
correspond to specific non-axisymmetric normal modes of the star. In
addition to the non-axisymmetric modes, the axisymmetric quasi-radial
mode could also be excited.

{\it Unstable Modes:} In both proto-neutron stars and hypermassive neutron
stars created in binary mergers, gravitational radiation can drive
several modes unstable, through the CFS mechanism (for a review, see
e.g.  Friedman \& Lockitch, 2001), provided the star is rotating
sufficiently rapidly. In relativistic stars, the $l=m=2$ $f$-mode
becomes unstable when $T/|W|>0.07$ for uniform rotation (Stergioulas \&
Friedman, 1998; Morsink, Stergioulas \& Blattnig, 1999) and at
somewhat larger $T/|W|$ for differentially rotating stars (Yoshida et
al. 2002).  The $l=m=2$ $r$-mode can become unstable at considerably
lower rotation rates (for a review see Kokkotas \& Andersson, 2001).
The suppression of the $r$-mode instability by the presence of
hyperons in the core (Jones, 2001; Linblom \& Owen, 2002) is not
expected to operate efficiently in rapidly rotating stars, since the
central density is probably too low to allow for hyperon formation.
Moreover, van Dalen \& Dieperink (2003) find the contribution of
hyperons to the bulk viscosity to be two orders of magnitude smaller
than previously estimated. If accreting neutron stars in Low Mass
X-Ray Binaries (LMXB, considered to be the progenitors of millisecond
pulsars) are shown to reach high masses of $\sim 1.8M_\odot$, then the
EOS could be too stiff to allow for hyperons in the core
(for recent observations that support a high mass for some millisecond
pulsars see Nice, Splaver \& Stairs, 2003). Both the $f$-mode and
the $r$-mode, when unstable, will grow on a timescale of several
seconds.

{\it Frequency of excited modes:} Even though in a typical nonrotating
neutron star the fundamental radial and quadrupole modes have
frequencies larger than about 1.5 kHz, in the rapidly rotating models
created in core collapse simulations, the frequency of these modes is
significantly reduced to well below 1 kHz (Dimmelmeier et al. 2002b),
which is within the sensitivity window of current gravitational-wave
detectors, such as VIRGO, GEO600, TAMA and LIGO.  The large reduction
in frequency is due to the fact that differential rotation allows much
lower central densities than uniform rotation. In the case of a binary
merger, the simulations by Shibata \& Ury\={u} (2002) have shown
quasi-periodic oscillations of a few kHz. The $r$-mode, if unstable,
would emit gravitational waves at a high frequency of 4/3 the
rotational frequency. These sources of gravitational waves are very
interesting for the proposed wide-band dual sphere detector (Cerdonio
et al. 2001). When the $f$-mode becomes unstable, it can have an
initial frequency of several hundred Hz, which then reduces to zero as
the star spins down and approaches the rotation rate at which the mode
becomes stable again. Therefore, the $f$-mode would ideally sweep
through the sensitivity window of current laser-interferometric
detectors, provided it can grow to a sufficiently large nonlinear
amplitude (Lai \& Shapiro, 1995).

{\it Damping of pulsations:} The stable pulsations excited in the
relativistic and Newtonian rotational core collapse simulations are
seen to be strongly damped.  In an isolated star, quasi-radial modes
are thought to be damped mainly by shear viscosity and gravitational
radiation (on a timescale of at least thousands of oscillation
periods). In principle, any pulsation of a magnetized star will also
be damped by the magnetic field (see e.g. McDermott et al. 1984;
Carroll et al. 1986). In the case of quasi-radial pulsations of
proto-neutron stars, however, we do not have specific estimates for
the corresponding damping time.  Once excited, the quasi-radial mode
would be an ideal long-term monochromatic source of gravitational
waves. However, the strong damping seen in the above simulations
reveals that a different mechanism operates in a proto-neutron star,
immediately after core bounce (Pons 2003).  The strong damping is due
to the presence of a high-entropy envelope, which surrounds the newly
created neutron star immediately after its birth.  The quasi-radial
oscillations penetrate into the envelope and are not properly
reflected by a sharp surface (as happens in the case of an isolated
star). The envelope thus absorbs much of the initial pulsation energy
on a dynamical timescale (a similar damping has been observed in
nonlinear evolutions of unstable relativistic stars, forced to migrate
to the stable branch of equilibrium models, see Font et al. 2002).  By
the time the proto-neutron star cools down (on a timescale of several
seconds) the initial pulsation amplitude has diminished.  Therefore,
the expected gravitational-wave signal from pulsations in core
collapse will be strongly damped. 

In order to model the expected signal, a detailed analysis of the
damping mechanism is required (taking into account various factors,
such as the EOS, the initial rotating stellar core etc.).  The
unstable $f$- and $r$-modes grow on a timescale of the order of
several seconds or longer, by which time the initial pulsations have
been dramatically damped. In the case of a binary merger, a
high-entropy envelope will also form.  Its extent will depend on the
mass ratio of the two stars -- an equal-mass irrotational binary could
create only a small envelope, which will not dampen the pulsations as
quickly as in a core collapse, while an unequal-mass binary merger
could result in a somewhat larger envelope being formed (see Shibata,
Taniguchi \& Ury\=u, 2003) and consequently in a stronger damping of
quasi-periodic pulsations.  In any case, the quasi-periodic
oscillations in a binary merger will be much more long-lived than in a
core collapse.

Pulsation modes of rapidly rotating stars, assuming uniform rotation,
have been computed by Yoshida and Eriguchi (1999, 2001), Font,
Stergioulas \& Kokkotas (2000), Font et al. (2001) and Stergioulas
\& Font (2001) in the Cowling approximation (in which spacetime
perturbations are neglected). In full general relativity, Stergioulas
\& Friedman (1998) and Morsink, Stergioulas \& Blattnig (1999)
computed the onset of the $l=m=2$ $f$-mode instability, while Font et
al. (2002) computed the frequencies of the two lowest-order
quasi-radial modes. For differentially rotating stars, the only
existing computation is by Yoshida et al. (2002), who obtained the
$l=m=2$ $f$-mode frequencies for a moderate strength of
differential rotation. Nonlinear effects of radial pulsations in
nonrotating relativistic stars have been studied by Sperhake,
Papadopoulos \& Andersson (2001) (see also Sperhake, 2002).

In the present paper, we study in detail two different sequences of
uniformly and differentially rotating relativistic polytropes, and
obtain the eigenfrequencies and eigenfunctions of several pulsations
modes in the Cowling approximation. For a sequence of fixed central
energy density the mode-frequencies are only weakly affected by the
rotation rate.  However, for a sequence of fixed rest mass, the mode
frequencies continuously decrease as the rotation rate increases (and
the central density decreases). An interesting result of our study is
that the fundamental quasi-radial mode splits into two different
modes, an effect which is more prominently seen in differentially
rotating models. Based on the frequency spectrum of nonrotating and
uniformly rotating stars, this split is not expected.  If it turns out
not to be really a new mode, it could be due to the fact that in the
Cowling approximation the energy and momentum conservation are
violated. For example, this violation leads to the appearance of an
unphysical ``fundamental'' dipole mode in simulations of nonrotating
stars (Font et al. 2001).  Another important outcome of our
investigation is that when studying the excitation of quasi-radial
pulsations in rapidly rotating models (which are near their
mass-shedding limit) we find that the pulsations are damped because of
mass-shedding in the equatorial region. This new damping mechanism
could set a severe limit on the saturation amplitude of unstable
modes.

The rest of the paper is organized as follows. In Section \ref{sec:2}
we outline our computational method, while in Section \ref{sec:3} we
describe the equilibrium properties of all initial models we subsequently
evolve.  In Section \ref{sec:4} the eigenfrequencies and
eigenfunctions of pulsating stars are presented while Section
\ref{sec:5} focuses on the splitting of the fundamental mode. Sections
\ref{sec:6} and \ref{sec:7} discuss the damping of pulsations due to
mass-shedding and the implications of our results on the detection of
gravitational waves, respectively. Finally, a summary and a discussion
of our results is presented in Section \ref{sec:8}.

\section{Outline of Computational method}
\label{sec:2}

We study the axisymmetric pulsations of rapidly rotating relativistic
stars by first constructing several sequences of uniformly and
differentially rotating equilibrium models, as described in detail in
Section \ref{sec:3}. The equilibrium models assume a relativistic
polytropic EOS of the form
\begin{eqnarray}
p &=& K \rho^{1+1/N}, \label{EOS1} \\
\varepsilon &=& \rho+Np,
\label{EOS2}
\end{eqnarray}
where $p$ is pressure, $\varepsilon$ is energy density, $\rho$ is
rest-mass density, $N$ is the polytropic index and $K$ is the
polytropic constant. Unless otherwise noted, we choose dimensionless
units for all physical quantities by setting $c=G=M_\odot=1$.

The nonlinear time-evolutions of perturbed equilibrium models are
carried out using the same numerical hydrodynamics code developed by
Font et al. (2001, 2002).  Suitable perturbations of selected
equilibrium variables are added to the equilibrium model, in order to
excite specific modes. In the absence of the true eigenfunction of a
given mode, each perturbation is chosen so as to mimic the angular
dependence of the eigenfunction of the corresponding mode of a
slowly-rotating Newtonian star. Usually, this ensures that the chosen
mode will dominate the time-evolution at least for the slower rotating
models.  However, since the perturbation is not exact, additional
pulsation modes will be excited, especially for rapidly rotating
models.  For the $l=0$ modes a density perturbation is used, of the
form
\begin{equation}
 \delta \rho = a \rho_c \sin \left(\pi\frac{ r}{r_s(\theta)}\right),
\label{drho}
\end{equation}
where $\rho_c$ is the central density of the star and $r_s(\theta)$ is
the coordinate radius of the surface of the star. The constant
$a$ is the amplitude of the perturbation, which we normally take
to be of the order of $1\%$. The $l=2$ modes are excited, by
perturbing the $\theta$-component of the four-velocity as follows:
\begin{equation}
 u_\theta= a \sin\left(\pi \frac{r}{r_s(\theta)}\right)
\sin\theta \cos\theta,
\end{equation}
(see Font et al. 2001 for more details).

The perturbed models are evolved in time with a two-dimensional
general-relativistic hydrodynamics code, in which we have implemented
a Godunov-type scheme based on the Marquina flux formula and the
3rd-order PPM reconstruction (see Font et al. 2000, 2001; see also
Font, 2003, for a review of Godunov-type schemes in general relativistic
hydrodynamics).  Keeping the spacetime fixed to the initial
equilibrium state during the evolution corresponds to the Cowling
approximation in perturbation theory. Below the mass-shedding limit
(see Friedman, Ipser \& Parker, 1986, for the precise definition of the
mass-shedding limit in the case of rapidly rotating relativistic stars),
we assume that the star remains isentropic by enforcing the EOS
(\ref{EOS1}),(\ref{EOS2}).  Near the mass-shedding limit, when shocks
form (see Section \ref{sec:6}), the adiabatic ideal fluid EOS is used
instead \begin{equation}
p=(\Gamma-1)\rho \epsilon, \label{if}
\end{equation}
where $\Gamma=1+1/N$ and $\epsilon$ is the specific internal
  energy.  Notice that at the initial time the isentropic equilibrium
  models constructed with the polytropic EOS (\ref{EOS1}),(\ref{EOS2}) are
  consistent with the ideal fluid EOS (\ref{if}).

The time series of the evolved perturbations are Fourier analyzed, and
the peaks in the corresponding spectra are identified with specific
pulsation modes, starting from the nonrotating member of the sequence,
where the pulsation frequencies are known from perturbation theory. As
the rotation rate increases, it becomes increasingly more difficult to
identify specific modes in the Fourier spectrum. For this reason, we
also extract the eigenfunction for each peak in the Fourier spectrum
and use it as an additional criterion to identify specific modes
(see Section \ref{sec:4}).

\section{Equilibrium models}
\label{sec:3}

Equilibrium models of rotating relativistic stars are constructed
using the numerical code {\tt rns} (Stergioulas \& Friedman, 1995)
which was extended to include differential rotation. Extensive tests
of the accuracy of the code in the case of uniform rotation can be
found in Nozawa et al. (1998) and in Stergioulas (2003). The metric
describing an axisymmetric relativistic star is assumed to have the
usual form
\begin{equation}
 ds^2 = -e^{\gamma+\rho} dt^2 + e^{\gamma-\rho} r^2 \sin^2
 \theta (d\phi - \omega dt)^2 + e^{2 \alpha} (dr^2 + r^2 d\theta^2),
\label{metric}
 \end{equation}
 where the metric functions $\gamma, \rho, \omega$ and $\alpha$ 
depend on the coordinates $r$ and $\theta$ only\footnote{The metric
 function $\rho$ in Eq. (\ref{metric}) should not be confused with
 the rest-mass density in Eq. (\ref{EOS1}).}. Axisymmetry enforces
 the specific angular momentum measured by the proper time of matter
 $j \equiv u^t u_\phi$ to be a function of the angular velocity
 $\Omega$ only. A different specific angular momentum
\begin{equation}
\tilde j = u_\phi \left( \varepsilon+p \over \rho \right ),
\end{equation}
is locally conserved during the phase of homologous collapse of a
rotating star. The Rayleigh criterion for local dynamical stability to
axisymmetric perturbations is
\begin{equation}
\frac{d\tilde j}{d \Omega} <0.
\end{equation}
The simplest common choice of the differential rotation law $j=j(\Omega)$ that
satisfies the Rayleigh stability criterion (see Komatsu, Eriguchi \&
Hachisu, 1989a,b) leads to an angular velocity distribution of the form
\begin{equation}
\Omega_c - \Omega=
\frac{1}{A^2} \left[ \frac{(\Omega-\omega)r^2 \sin^2 \theta e^{-2
      \rho}} {1-(\Omega-\omega)^2 r^2 \sin^2 \theta e^{-2 \rho}}
\right],
 \label{difrotlaw}
 \end{equation}
where $\Omega_c$ is the angular velocity on the rotational axis
and  $A$ is a parameter with units of length, that determines
the length scale over which the angular velocity varies inside
the star (in the limit of $A \rightarrow \infty$, uniform rotation
is recovered).

When constructing sequences of fixed rest mass, the radius of the star
can vary by more than a factor of two. If one would characterize the
sequence by a fixed value of the parameter $A$, then different models
would correspond to different degrees of differential rotation.  In
order to ascribe the same degree of differential rotation to all
models along a sequence, we follow Baumgarte, Shapiro \& Shibata
(2000) in normalizing the parameter $A$ by the radius of the star.
Thus, we define
\begin{equation}
{\hat A}=A/r_e,
\end{equation}
where $\hat A$ is now a dimensionless parameter and $r_e$ is the
equatorial coordinate radius of the star. Notice that in the Newtonian
limit, ${\hat A}$ is the radius of the cylinder, as a fraction of the
radius of the star, where the angular velocity falls to one half of
the central angular velocity.

Our focus is on the effect of rotation on pulsation modes. Hence, we
do not survey a broad range of suggested high-density equations of
state, but rather choose a polytropic EOS with $N=1$ and $K=100$,
respectively. This choice corresponds to models of neutron stars
having mass and radius similar to those constructed with a realistic
EOS of average stiffness.  We focus attention on two different
sequences of differentially rotating models (sequences A and B) and
their uniformly rotating counterparts (sequences AU and BU). The
equilibrium properties of all models are displayed in Table
\ref{tabeq}. All sequences terminate at the same nonrotating model
(thus, models A0, AU0, B0 and BU0 all coincide).


\begin{table*}
\begin{minipage}{140mm}
\begin{center}
\caption{Properties of the four sequences of equilibrium models (A is 
  a sequence of fixed rest mass $M_0=1.506 M_\odot$ with $\hat A=1$,
  AU is the corresponding sequence of uniformly rotating models, B is
  a sequence of fixed central rest mass density $\rho_c=1.28 \times
  10^{-3}$ with $\hat A=1$, and BU is the corresponding sequence of
  uniformly rotating models).  All models are relativistic polytropes
  with $N=1$ and $K=100$. The definitions of the various quantities
  are given in the main text. Notice that all quantities are in
  dimensionless units with $c=G=M_\odot=1$.}
\begin{tabular}{*{9}{c}}
\\
\hline
model  &  $\varepsilon_c$  &     $M$  &  $R$  &  $r_e$  &  $r_p/r_e$  &  
$\Omega_c$  &  $\Omega_e$  &  $T/|W|$   \\[0.5ex]
       &  $(\times 10^{-3})$ &       &       &   &   &  $(\times 10^{-2})$  &  
$(\times 10^{-2})$   &    \\[0.5ex]
\hline
A0  &  1.444  &  1.400  &   9.59  &  8.13  &  1.0    &  0.0    &  0.0    &   0.0    
\\[0.5ex]
A1  &  1.300  &  1.405  &  10.01  &  8.54  &  0.930  &  0.202  &  0.076  &   0.018  
\\[0.5ex]
A2  &  1.187  &  1.408  &  10.40  &  8.92  &  0.875  &  0.258  &  0.098  &   0.033  
\\[0.5ex]
A3  &  1.074  &  1.410  &  10.84  &  9.35  &  0.820  &  0.294  &  0.113  &   0.049  
\\[0.5ex]
A4  &  0.961  &  1.413  &  11.37  &  9.87  &  0.762  &  0.319  &  0.123  &   0.066  
\\[0.5ex]
A5  &  0.848  &  1.418  &  12.01  & 10.49  &  0.703  &  0.334  &  0.130  &   0.086  
\\[0.5ex]
A6  &  0.735  &  1.422  &  12.78  & 11.25  &  0.643  &  0.338  &  0.134  &  0.107  
\\[0.5ex]
A7  &  0.622  &  1.427  &  13.75  & 12.21  &  0.579  &  0.334  &  0.134  &  0.131  
\\[0.5ex]
A8  &  0.509  &  1.433  &  15.01  & 13.45  &  0.513  &  0.320  &  0.130  &  0.158  
\\[0.5ex]
A9 &  0.396  &  1.439  &  16.70  & 15.13  &  0.444  &  0.295  &  0.122  &   0.189  
\\[0.5ex]
A10 &  0.283  &  1.447  &  19.03  & 17.44  &  0.370  &  0.260  &  0.110  &  0.223  
\\[0.5ex]
A11 &  0.170  &  1.456  &  21.92  & 20.30  &  0.294  &  0.218  &  0.094  &  0.260  
\\[0.5ex]
\hline
AU0  &  1.444  &  1.400  &   9.59  &  8.13  &  1.0    &  0.0    &  0.0    &   0.0    
\\[0.5ex]
AU1 &  1.300  & 1.404   &  10.19  &  8.71  &  0.919  &  1.293   &  1.293  &  0.020   
\\[0.5ex]
AU2 &  1.187  & 1.407   &  10.79  &  9.30  &  0.852  &  1.656   &  1.656  &  0.037   
\\[0.5ex]
AU3 &  1.074  & 1.411   &  11.56  & 10.06  &  0.780  &  1.888   &  1.888  &  0.055   
\\[0.5ex]
AU4 &  0.961  & 1.415   &  12.65  & 11.14  &  0.698  &  2.029   &  2.029  &  0.076   
\\[0.5ex]
AU5 &  0.863  & 1.420   &  14.94  & 13.43  &  0.575  &  2.084   &  2.084  &  0.095   
\\[0.5ex]
\hline
B0  &  1.444  &  1.400  &   9.59  &  8.13  &  1.0    &  0.0    &  0.0    &   0.0    
\\[0.5ex]
B1  &   1.444  & 1.437   &   9.75  &  8.24  &  0.950  &  0.180  &  0.067  &  0.013   
\\[0.5ex]
B2  &   1.444  & 1.478   &   9.92  &  8.36  &  0.900  &  0.257  &  0.094  &  0.026   
\\[0.5ex]
B3  &   1.444  & 1.525   &  10.11  &  8.49  &  0.849  &  0.319  &  0.116  &  0.040   
\\[0.5ex]
B4  &  1.444   & 1.578   &  10.31  &  8.63  &  0.800  &  0.373  &  0.134  &  0.055   
\\[0.5ex]
B5  &   1.444  & 1.640   &  10.53  &  8.77  &  0.750  &  0.423  &  0.150  &  0.071   
\\[0.5ex]
B6  &   1.444  & 1.713   &  10.76  &  8.91  &  0.700  &  0.471  &  0.165  &  0.087   
\\[0.5ex]
B7  &   1.444  & 1.798   &  11.01  &  9.05  &  0.650  &  0.519  &  0.179  & 0.105   
\\[0.5ex]
B8  &   1.444  & 1.899   &  11.26  &  9.17  &  0.600  &  0.568  &  0.192  & 0.124   
\\[0.5ex]
B9  &   1.444  & 2.020   &  11.50  &  9.26  &  0.550  &  0.623  &  0.205  & 0.144   
\\[0.5ex]
B10 &   1.444  & 2.167   &  11.71  &  9.27  &  0.500  &  0.689  &  0.219  & 0.165   
\\[0.5ex]
B11 &   1.444  & 2.341   &  11.80  &  9.13  &  0.450  &  0.777  &  0.236  & 0.187   
\\[0.5ex]
B12 &   1.444  & 2.532   &  11.64  &  8.72  &  0.400  &  0.912  &  0.258  & 0.207   
\\[0.5ex]
\hline
BU0  &  1.444  &  1.400  &  9.59   &  8.13  &  1.00  &  0.0    &  0.0    &   0.0 
\\[0.5ex]
BU1  &  1.444  &  1.432  &  9.83   &  8.33  &  0.95  &  1.075  &  1.075  &  0.012 
\\[0.5ex]
BU2  &  1.444  &  1.466  &  10.11  &  8.58  &  0.90  &  1.509  &  1.509  &  0.024 
\\[0.5ex]
BU3  &  1.444  &  1.503  &  10.42  &  8.82  &  0.85  &  1.829  &  1.829  &  0.037 
\\[0.5ex]
BU4  &  1.444  &  1.543  &  10.78  &  9.13  &  0.80  &  2.084  &  2.084  &  0.050 
\\[0.5ex]
BU5  &  1.444  &  1.585  &  11.20  &  9.50  &  0.75  &  2.290  &  2.290  &  0.062 
\\[0.5ex]
BU6  &  1.444  &  1.627  &  11.69  &  9.95  &  0.70  &  2.452  &  2.452  &  0.074 
\\[0.5ex]
BU7  &  1.444  &  1.666  &  12.30  &  10.51  &  0.65  &  2.569  &  2.569  &  0.084 
\\[0.5ex]
BU8  &  1.444  &  1.692  &  13.07  &  11.26  &  0.60  &  2.633  &  2.633  &  0.091 
\\[0.5ex]
BU9  &  1.444  &  1.695  &  13.44  &  11.63  &  0.58  &  2.642  &  2.642  &  0.092 
\\[0.5ex]
\hline
\label{tabeq}
\end{tabular}
\end{center}
\end{minipage}
\end{table*}

\begin{figure}
  \centerline{\psfig{file=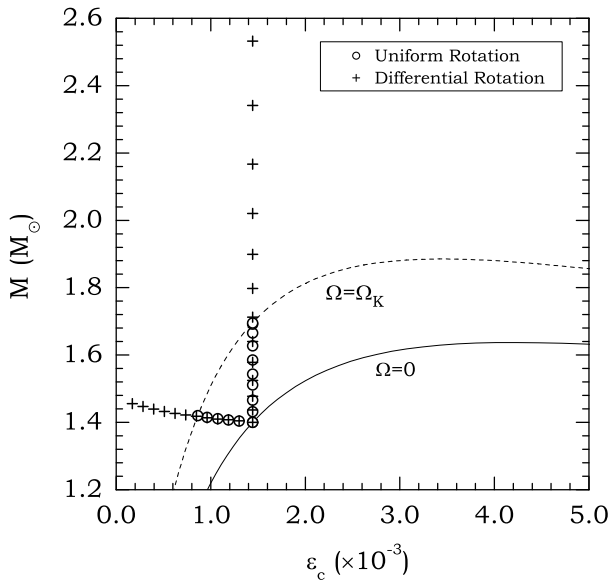,width=8cm}}
  \caption{Mass vs. central energy density plot of all models in
    sequences A, AU, B and BU. Open circles correspond to
    uniformly rotating models, while the crosses correspond to
    differentially rotating models.  The two nearly horizontal
  sequences are A and AU, while the two vertical sequences are B
  and BU.  Also shown are the sequence of nonrotating models (solid
  line), and the sequence of models rotating at the mass-shedding
  limit for uniform rotation (dashed line).  Differential rotation
  allows equilibrium models well beyond the region allowed for
  uniformly rotating models.}
\label{fig:massvsdensity}
\end{figure}

The differentially rotating sequence A and its corresponding uniformly
rotating sequence AU are characterized by a fixed rest mass $M_0=1.506
M_\odot$.  Along sequence A, the degree of differential rotation is
held fixed at $\hat A=1$. The values of $M_0$ and $\hat A$ are chosen
in order to represent a newly-born, differentially rotating neutron
star.  The angular velocity at the equator is roughly 1/3 to 1/2 of
the central angular velocity, which is similar to the degree of
differential rotation obtained in typical core collapse simulations
(see Villain et al. 2003). The fastest rotating model in sequence A
has a polar to equatorial coordinate axes ratio of only
$r_p/r_e=0.294$, a ratio $T/|W|=0.26$ and rotates close to, but still
below, the mass-shedding limit.  The central density is nearly an
order of magnitude smaller than the corresponding nonrotating model,
while the circumferential radius is more than twice as
large. The uniformly rotating sequence AU only reaches an axes ratio
of 0.575, a ratio $T/|W|=0.095$, half the central
density, and a 50\% larger radius than the corresponding
nonrotating model. Model AU5 is at the mass-shedding limit.

 \begin{figure}
\includegraphics[width=8cm,clip]{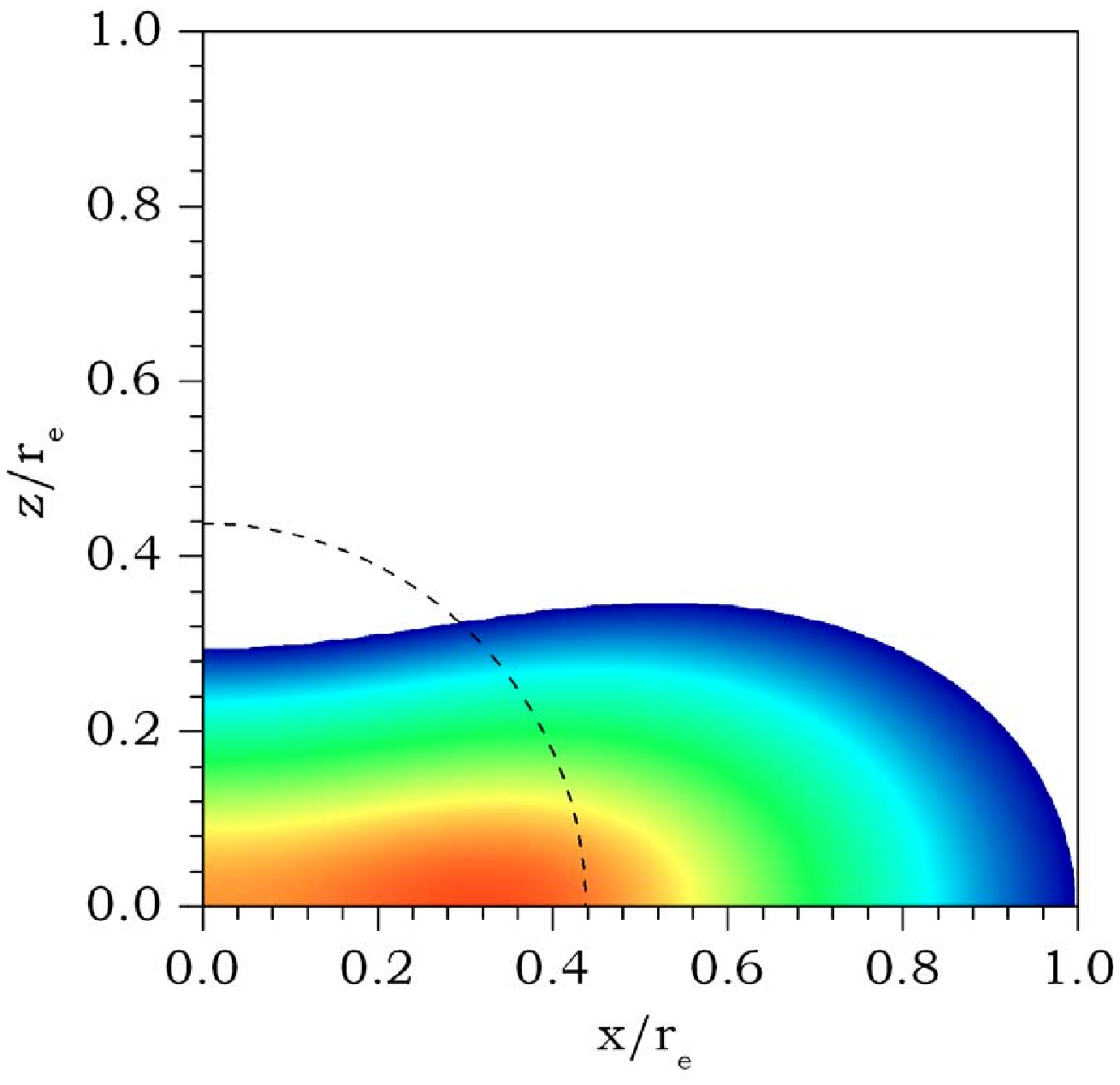}
   \caption{Density stratification in the fastest differentially
    rotating model of the fixed rest-mass sequence A. The maximum
    density appears off-center, at $x/r_e=0.32$. In comparison, the
    shape of the nonrotating star of same rest mass is shown (dashed
    line), scaled by the equatorial radius of the rotating model.}
 \label{fig:contours}
 \end{figure}
 
 On the other hand, the differentially rotating sequence B and its
 corresponding uniformly rotating sequence BU are characterized by a
 fixed central density $\rho_{c}=1.28 \times 10^{-3}$ and fixed $\hat
 A=1$.  Its fastest rotating member has rest mass of $M_0=2.79
 M_\odot$ and a gravitational mass of $2.53 M_\odot$, corresponding
 roughly to a hypermassive neutron star created in a binary neutron
 star merger.  The degree of differential rotation is in rough
 agreement with simulations by Shibata \& Ury\={u} (2000, 2002). The
 axes ratio for the fastest rotating model is 0.4 (the model being
 still below, but close to, the mass-shedding limit).  Since all
 models in the sequence are compact, the radius $R$ only increases by
 21\%.  The corresponding uniformly rotating sequence only reaches an
 axes ratio of 0.58 at the mass-shedding limit (BU9) with an increase
 in radius $R$ by 40\%. Thus, we see that when considering a sequence
 of fixed central density, the uniformly rotating models attain a
 larger equatorial radius than differentially rotating models, which
 tend to expand out of the equatorial plane, becoming torus-like.
 
 Fig. \ref{fig:massvsdensity} shows all constructed equilibrium models
 in a mass vs. central energy density plot. The two nearly horizontal
 sequences are A and AU, while the two vertical sequences are B and
 BU.  Open circles correspond to uniformly rotating models, while
 the crosses correspond to differentially rotating models.  Also shown are
 the sequence of nonrotating models (solid line) and the sequence of
 models rotating at the mass-shedding limit for uniform rotation
 (dashed line).  Differential rotation allows equilibrium models well
 beyond the region allowed for uniformly rotating models (see also
Baumgarte, Shapiro \& Shibata, 2000, where such configurations are called
{\it hypermassive}).

In Fig. \ref{fig:contours} the density stratification for the fastest
rotating model of sequence A (A11) is shown. In the figure, Cartesian
coordinates, $x=r\sin\theta, z=r\cos\theta$ are used, scaled by the
equatorial coordinate radius $r_e$. The maximum density is attained
off-center.  The dashed line shows the spherical surface of the
nonrotating model with the same rest mass (A0), scaled by the equatorial
coordinate radius of the rotating model A11. It becomes evident that the
large rotation rate of model A11 causes the equatorial radius to increase
by more than a factor of two.

Finally, we note that apart from the above two sequences, we have also
investigated the effect of the degree of differential rotation on the
axisymmetric pulsations, by constructing various sequences that differ
only in the value of $\hat A$. 

\begin{table}
\begin{center}
\caption{Frequencies of the fundamental quasi-radial ($l=0$) mode, $F$, 
  its first overtone, $H_1$, the fundamental quadrupole ($l=2$) mode,
  $^{2}f$ and its first overtone, $^{2}p_1$, for the sequence of
  uniformly rotating models AU.}
\begin{tabular}{|c|c|c|c|c|}
  \hline
  model & $F$ &    $H_1$ & $^{2}f$ & $^{2}p_1 $ \\
        & kHz &    kHz & kHz & kHz \\
  \hline
AU0  &   2.706  &    4.547  &     1.846  &     4.100 \\
AU1  &   2.526  &     4.246  &    1.800  &     3.862 \\
AU2  &   2.403  &     4.090  &    1.744  &     3.592 \\
AU3  &   2.277  &     3.937  &    1.663  &     3.265 \\
AU4  &   2.141  &     3.795  &    1.547  &     2.847 \\
AU5  &   1.960  &     3.647  &    1.330  &     2.560 \\
  \hline
\protect\label{tabAU}
\end{tabular}
\end{center}
\end{table}

\begin{table}
\begin{center}
\caption{Same as Table \protect\ref{tabAU}, but for the sequence of 
  differentially rotating models A. Also shown are the frequencies of an
  additional fundamental mode $F_{\rm II}$ that appears mainly in
  differentially rotating models (at least in the Cowling
  approximation). }
\begin{tabular}{|c|c|c|c|c|c|}
  \hline
  model & $F_{\rm II}$ &   $F$   &  $H_1$ & $^{2}f$ & $^{2}p_1 $ \\
        & kHz &  kHz  & kHz & kHz & kHz \\
  \hline
A0  &   2.706  &   2.706  &     4.547  &     1.846  &     4.100 \\
A1  &   2.485  &   2.561  &     4.310  &     1.822  &     3.961 \\
A2  &   2.361  &   2.480  &     4.163  &     1.780  &     3.822 \\
A3  &   2.243  &   2.386  &     4.029  &     1.738  &     3.642 \\
A4  &   2.142  &   2.295  &     3.900  &     1.677  &     3.442 \\
A5  &   2.039  &   2.201  &     3.748  &     1.598  &     3.211 \\
A6  &   1.921  &   2.101  &     3.563  &     1.510  &     2.973 \\
A7  &   1.779  &   1.982  &     3.329  &     1.403  &     2.711 \\
A8  &   1.609  &   1.846  &     3.120  &     1.274  &     2.494 \\
A9  &   1.424  &   1.667  &     2.857  &     1.132  &     2.185 \\
A10 &   1.223  &   1.422  &     2.503  &     0.966  &     1.905 \\
A11 &   1.044  &   1.220  &     2.174  &     0.820  &     1.658 \\
  \hline
\end{tabular}
\label{tabA}
\end{center}
\end{table}

\section{Eigenfrequencies and eigenfunctions}
\label{sec:4}

Although we use a nonlinear evolution code to study pulsations of
rotating stars, we restrict attention to small-amplitude pulsations
(small in the sense that e.g. $\delta \rho / \rho \sim 10^{-2}$).
Therefore, the time-evolution of a perturbed star can still be viewed
(to a good approximation) as a superposition of linear normal modes.
When obtaining the Fourier transform of the time-evolution of several
variables, we verify that the various modes that are excited have indeed
discrete frequencies (same frequency at any point inside the star in
the coordinate frame).  Thus, even though the evolution is nonlinear,
the amplitude is small enough to justify the use of the terms
``eigenfrequency'' and ``eigenfunction'' for the various pulsation
modes. In order to compute the real part of the eigenfrequency of a
pulsation mode we Fourier-transform the time series of the evolution
of a suitable physical variable (the density for the $l=0$ modes and
$v_\theta$ for the $l=2$ modes). Instead of examining the Fourier
spectra at a few specific points inside the star, we integrate the
amplitude of the Fourier transform along a coordinate line, e.g. for
the $l=0$ modes we examine the integrated Fourier amplitude along
$\theta=\pi/2$ (equatorial plane), while for the $l=2$ modes the
integrated Fourier amplitude along a line of $\theta=\pi/4$ is used.

\begin{table}
\begin{center}
\caption{Same as Table \ref{tabAU}, but for the sequence of uniformly rotating 
models BU.}
\begin{tabular}{|c|c|c|c|c|}
  \hline
  model & $F$ &    $H_1$ & $^{2}f$ & $^{2}p_1 $ \\
        & kHz &    kHz & kHz & kHz \\
  \hline
BU0  &   2.706  &      4.547  &    1.846  &     4.100 \\
BU1  &   2.657  &      4.467  &    1.855  &     4.040 \\
BU2  &   2.619  &      4.409  &    1.860  &     3.944 \\
BU3  &   2.579  &      4.385  &    1.857  &     3.814 \\
BU4  &   2.535  &      4.371  &    1.844  &     3.645 \\
BU5  &   2.495  &      4.356  &    1.815  &     3.456 \\
BU6  &   2.456  &      4.357  &    1.762  &     3.244 \\
BU7  &   2.417  &      4.337  &    1.686  &     3.010 \\
BU8  &   2.328  &      4.300  &    1.588  &     2.710 \\
BU9  &   2.313  &      4.280  &    1.558  &     2.642 \\
  \hline
\end{tabular}
\end{center}
\label{tabBU}
\end{table}

\begin{table}
\begin{center}
\caption{Same as Table \ref{tabA}, but for the sequence of 
  differentially rotating models B.}
\begin{tabular}{|c|c|c|c|c|c|}
  \hline
  model & $F_{\rm II}$ &   $F$   &  $H_1$ & $^{2}f$ & $^{2}p_1 $ \\
        & kHz &  kHz  & kHz & kHz & kHz \\
  \hline
B0  &   2.706  &    2.706  &    4.547  &    1.846  &     4.100 \\
B1  &   2.627  &    2.658  &    4.446  &    1.880  &     4.102 \\
B2  &   2.561  &    2.637  &    4.421  &    1.900  &     4.090 \\
B3  &   2.525  &    2.632  &    4.405  &    1.913  &     4.045 \\
B4  &   2.506  &    2.632  &    4.403  &    1.924  &     3.983 \\
B5  &   2.487  &    2.632  &    4.422  &    1.929  &     3.907 \\
B6  &   2.459  &    2.633  &    4.436  &    1.925  &     3.828 \\
B7  &   2.423  &    2.635  &    4.447  &    1.909  &     3.721 \\
B8  &   2.394  &    2.646  &    4.444  &    1.890  &     3.632 \\
B9  &   2.360  &    2.653  &    4.413  &    1.867  &     3.567 \\
B10 &   2.330  &    2.662  &    4.360  &    1.842  &     3.500 \\
B11 &   2.318  &    2.678  &    4.300  &    1.832  &     3.470 \\
B12 &   2.328  &    2.722  &    4.240  &    1.830  &     3.473 \\
  \hline
\end{tabular}
\label{tabB}
\end{center}
\end{table}

\begin{figure}
\includegraphics[width=8cm,clip]{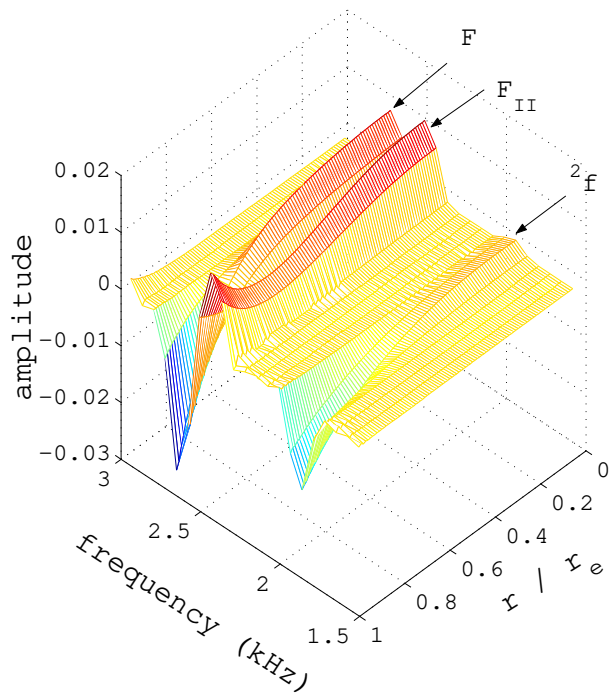}
\caption{Amplitude of the Fourier transform of the time-evolution of the 
  density in the equatorial plane, for model B7, after applying an
  $l=0$ perturbation. Several excited pulsation modes can be
  identified. For a particular mode, the amplitude of the Fourier
  transform correlates with its eigenfunction.}
\label{fig:eigen}
\end{figure}

Since the trial-eigenfunction used for exciting the pulsations does
not correspond exactly to a particular mode, several additional modes
are excited, apart from the main mode one wishes to study. This is
particularly true for very rapidly rotating models, where rotational
coupling effects are significant and higher-order coupling terms in
the mode-eigenfunctions become comparable to the dominant term. Thus,
in order to identify specific modes for several models along a
sequence, one has to begin with pulsations of a nonrotating star (for
which the modes are identified by comparison to results obtained with
linear perturbation theory, see Font et al. 2000, 2001,
2002)\footnote{For a review of pulsations of nonrotating relativistic
  stars see Kokkotas \& Schmidt (1999).} and gradually identify modes
for more rapidly rotating stars by comparing the peaks in a Fourier
transform to the corresponding peaks in a more slowly rotating model.
Close to the mass-shedding limit, avoided crossings between low-order
and high-order modes can complicate the picture and lead to
erroneous identifications.  For this reason, we do not only rely on
the Fourier transforms at a few points inside the star, but
reconstruct the whole two-dimensional eigenfunction of each mode,
using Fourier transforms at every point inside the star. At the
eigenfrequency of a specific mode, the amplitude of the Fourier
transform correlates with its eigenfunction. A change in sign in the
eigenfunction corresponds to both the real and imaginary part of the
Fourier transform going through zero. Comparing the eigenfunctions
corresponding to different peaks in a Fourier transform, allows for an
unambiguous identification of specific mode sequences. 

A example of the eigenfunction extraction using Fourier transforms is
shown in Figure \ref{fig:eigen}, which displays the amplitude of the
Fourier transform of the time-evolution of the density, after an $l=0$
perturbation was applied to model B7. In the range of frequencies
shown in Figure \ref{fig:eigen}, the amplitude of the Fourier
transform clearly correlates with the eigenfunctions of the
fundamental $F$- and $^2f$-modes (the additional mode, $F_{\rm
  II}$, is discussed in Section \ref{sec:5}).  Notice that in
displaying the eigenfunctions we use the amplitude of the Fourier
transform, multiplied by the sign of its real part.

When comparing the eigenfrequencies between models that differ only in
the degree of differential rotation, we see that a moderate degree of
differential rotation has some (but not dramatic) effect on the
eigenfrequencies.  When $\hat A$ is decreased significantly below
$\sim 1$, which leads to a ``strongly'' differentially rotating model,
most of the angular momentum of the star is concentrated in a narrow
region around the rotational axis. Outside this region the star is
only slowly rotating and the eigenfrequencies of its pulsations become
again similar to those of a nonrotating model. Thus, the effect of
differential rotation on the eigenfrequencies becomes strongest, not
for very small values of $\hat A$ (strong differential rotation), but
for values that correspond to stars with only a moderate degree of
differential rotation. 

\subsection{Fixed rest mass sequences}
\label{fixed}

It is well-known that the frequencies of the fundamental $l=0$ and $l=2$
polar modes of oscillation depend mainly on the central density of a
star, or, equivalently, on the compactness $M/R$ (see e.g. Hartle \&
Friedman 1975).  This is particularly true for the axisymmetric
($m=0$) modes.  The sequences of fixed rest mass $M_0=1.506 M_\odot$
start with a nonrotating model with compactness $M/R = 0.15$. Rotation
increases the radius and decreases the central density. The uniformly
rotating sequence AU terminates at the mass-shedding limit, with a
compactness of $M/R=0.095$. The differentially rotating sequence can
reach higher rotation rates and terminates near the mass-shedding
limit with a compactness of $M/R=0.066$. Based on the significant
decrease of the compactness along the fixed-rest-mass sequences, we
expect a corresponding decrease in the frequencies of the fundamental
modes (and a similar tendency for the first overtones).

\begin{figure}
  \centerline{\psfig{file=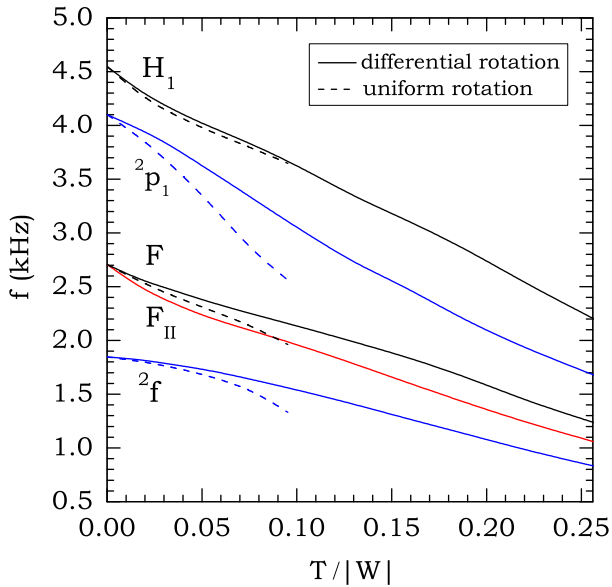,width=8cm}}
  \caption{Eigenfrequencies of various modes for sequences A and AU.
    The lower of the two lines that start at the $F$-mode frequency
    for the nonrotating models is an additional mode, $F_{\rm II}$,
    appearing mainly in differentially rotating stars (at least in the
    Cowling approximation).} 
\label{fig:freqsA}
\end{figure}

Table \ref{tabAU} displays the computed eigenfrequencies for the
fundamental and first overtone of the $l=0$ and $l=2$ modes for the
sequence AU. Rotation reduces the frequency of the fundamental
quasi-radial mode from 2.71 kHz to 1.96 kHz with corresponding changes
in the frequencies of the other modes. Figure \ref{fig:freqsA}
displays the variation of the mode-frequencies with increasing $T/|W|$
along sequence AU (dashed lines). The rate of decrease in frequency
for the fundamental quasi-radial modes, with increasing rotation rate
becomes larger as the mass-shedding limit is approached (due to a cusp
forming in the equatorial region), while the decrease in frequency for
the other modes is a nearly linear function of $T/|W|$.

Table \ref{tabA} reports the corresponding eigenfrequencies for the
differentially rotating sequence A, which are also shown in Figure
\ref{fig:freqsA} (solid lines). The frequencies of all modes decrease
nearly linearly with increasing $T/|W|$. Due to the differential
rotation, the outer layers of the star rotate slower and the
equatorial radius is smaller than a uniformly rotating model of same
$T/|W|$. This leads to a smaller sound-crossing time and correspondingly
higher fundamental mode frequencies for the differentially rotating
models.  This explains why the lines in Fig. \ref{fig:freqsA}
corresponding to the fundamental modes of sequence A have smaller
slopes than those corresponding to sequence AU.

For the fastest rotating model of sequence A, the fundamental
quasi-radial mode has a frequency of only 1.22 kHz, with the
fundamental quadrupole mode having a frequency of 0.82 kHz.  It should
be emphasized that the above frequencies are computed in the Cowling
approximation, which leads to higher values than the actual
frequencies (see Yoshida \& Kojima, 1997). Therefore, the actual
frequencies of the fundamental quasi-radial modes should be in the
range of $40-45$\% smaller than those computed here. The actual
frequencies of the fundamental quadrupole mode should be roughly
$15-20$\% smaller than our values.  For the nonrotating model of
sequences A and AU, the actual frequencies of the fundamental $l=0$
and $l=2$ modes are 1.44 kHz and 1.58 kHz, respectively (Font et al.
2002).  Thus, the fundamental $l=0$ mode has a frequency only somewhat
smaller than the fundamental $l=2$ mode (while in the Cowling
approximation it has a frequency significantly larger than the
fundamental $l=2$ mode).

In Font et al. (2002) it was found that, in the case of uniform
rotation, both the actual frequency and the frequency in the Cowling
approximation of the fundamental quasi-radial mode decrease in a
similar way as the rotation rate increases. This leads to the following
empirical relation between the actual frequency and the frequency in
the Cowling approximation.
\begin{equation}
f_\Omega=f^{\rm{(C)}}_\Omega+ \left(f_0
-f^{\rm{(C)}}_0 \right),
\end{equation}
where the superscript (C) denotes the Cowling approximation,
$f_\Omega$ is the frequency of the fundamental quasi-radial mode for a
rotating star with angular velocity $\Omega$, and $f_0$ is the
corresponding frequency in a nonrotating star. This empirical relation
was found to be accurate to within $2 \%$ at all rotation rates, even
near the mass-shedding limit. Since the frequencies of differentially
rotating stars are not much different with respect to those of
uniformly rotating stars with the same oblateness, the above relation
should approximately hold for differentially-rotating models as well.

Based on Yoshida \& Kojima (1997) and Yoshida \& Eriguchi (2001) we
can estimate the error in the fundamental mode frequencies, due to the
Cowling approximation. For the fastest rotating model of sequence A,
with $M/R \sim 0.07$, we estimate that the actual fundamental $l=0$
and $l=2$ frequencies are 63\% and 79\% of our results in the Cowling
approximation, respectively (actual frequency meaning the frequency
without the assumption of the Cowling approximation). Taking into
account the nearly linear scaling of the frequencies with increasing
rotation rate in Figure \ref{fig:freqsA}, we construct the following
empirical relation
\begin{equation}
f({\rm kHz})\approx 1.44-2.59 \frac{T}{|W|},
\end{equation}
for the actual frequency of the fundamental quasi-radial mode and
\begin{equation}
f({\rm kHz})\approx 1.58-3.69 \frac{T}{|W|},
\end{equation}
for the actual frequency of the fundamental quadrupole mode for the
differentially rotating models of sequence A. We estimate the uncertainty
of the above empirical relations to be on the order of a few percent. 

\begin{figure}
\centerline{\psfig{file=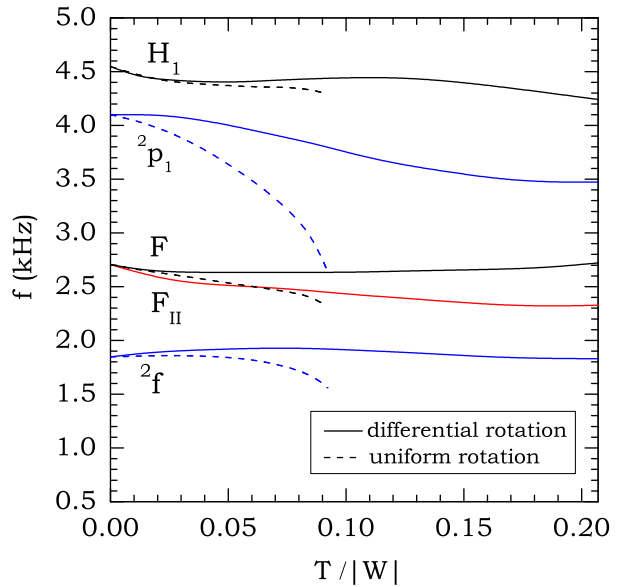,width=8cm}} 
\caption{Same as Figure 
\ref{fig:freqsA}, but for the sequences B
and BU.} \label{fig:freqsB}
\end{figure}

According to the above empirical relations, for rapidly rotating
proto-neutron stars with $T/|W|$ in the range $0.14 - 0.26$, the
frequency of the two fundamental modes will be in a range of roughly
$0.65-1.1$ kHz. This range of frequencies agrees well with the
frequencies of gravitational waves observed in the rotating core
collapse simulations by e.g. Dimmelmeier et al. (2001, 2002a, 2002b),
which confirms the validity of our chosen equilibrium models and
pulsation modes to model the gravitational waves produced in the above
simulations\footnote{Here we only refer to the regular collapse cases
  in Dimmelmeier et al. and not to multiple-bounce cases. Notice that,
  even though our polytropic index is different than in Dimmelmeier et
  al., our choice of the polytropic constant leads to models of
  similar compactness, for the same mass.}.

The additional fundamental frequency $F_{\rm II}$ in Tables \ref{tabAU}
and \ref{tabA} and in Figure \ref{fig:freqsA} is discussed in
Section \ref{sec:5}.

\subsection{Fixed central density sequence}

Another sequence of models with differential rotation that has been
analyzed is sequence B, consisting of twelve models with the same
central energy density $\epsilon_c=1.28 \times 10^{-3}$ but with axes
ratios ranging from 1.0 to 0.40. The most rapidly rotating model
corresponds to a hypermassive neutron star, such as those created
temporarily in a binary neutron star merger. The corresponding
sequence of uniformly rotating models, BU, of same fixed central
energy density, comprises models with axes ratios from 1.0 to 0.58. The
computed frequencies of the $l=0,2$ fundamental modes and their first
overtones are shown in Figure \ref{fig:freqsB} and in Tables
\ref{tabBU} and \ref{tabB} (the additional fundamental mode
$F_{\rm II}$ is discussed in Section \ref{sec:5}).

\begin{figure}
\includegraphics[width=8cm,clip]{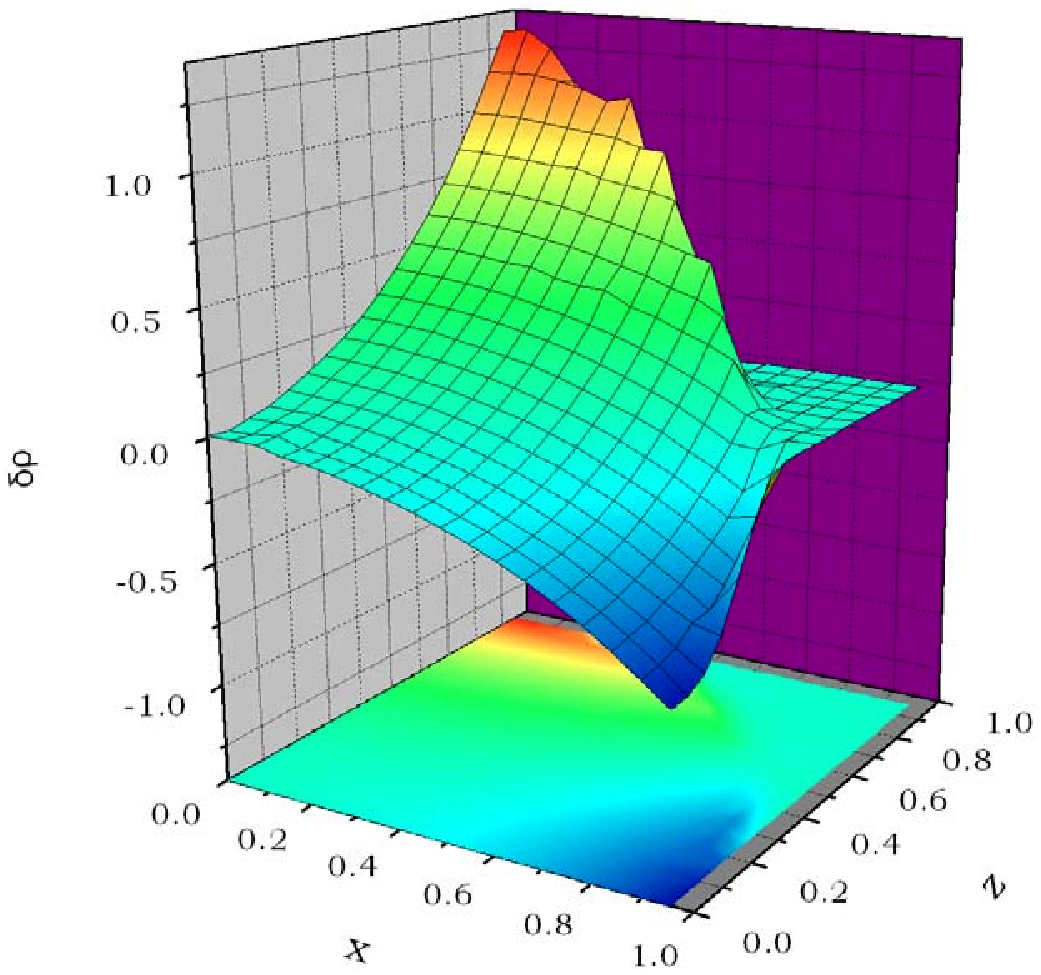}
\caption{Two-dimensional eigenfunction of the
  density perturbation, corresponding to the $l=2$ fundamental mode
  for the nonrotating model B0. The eigenfunction is shown in
  Cartesian coordinates with $x$ and $z$ being scaled by the
  equatorial coordinate radius $r_e$, while $\delta \rho$ is in
  arbitrary units.}
\label{fig:l2B0}
\end{figure}

The frequencies of the fundamental $l=0$ and $l=2$ modes changes by as
much as 16\% along the sequence BU.  In comparison, for sequence B the
eigenfrequencies for both the $l=0$ and $l=2$ fundamental modes remain
relatively unaffected by the rotation rate, changing by less than 3\%
and 5\%, respectively, compared to their values in the nonrotating
limit.  This leads to the conclusion that a moderate amount of
differential rotation, of the order of $\hat A=1$, makes the
fundamental frequencies relatively insensitive to rotation and
depending mainly on the central energy density of the star. This
finding should simplify attempts to extract information about the
physical properties of neutron stars in binary mergers, when such
events become observable through their gravitational-wave emission.

\begin{figure}
\includegraphics[width=8cm,clip]{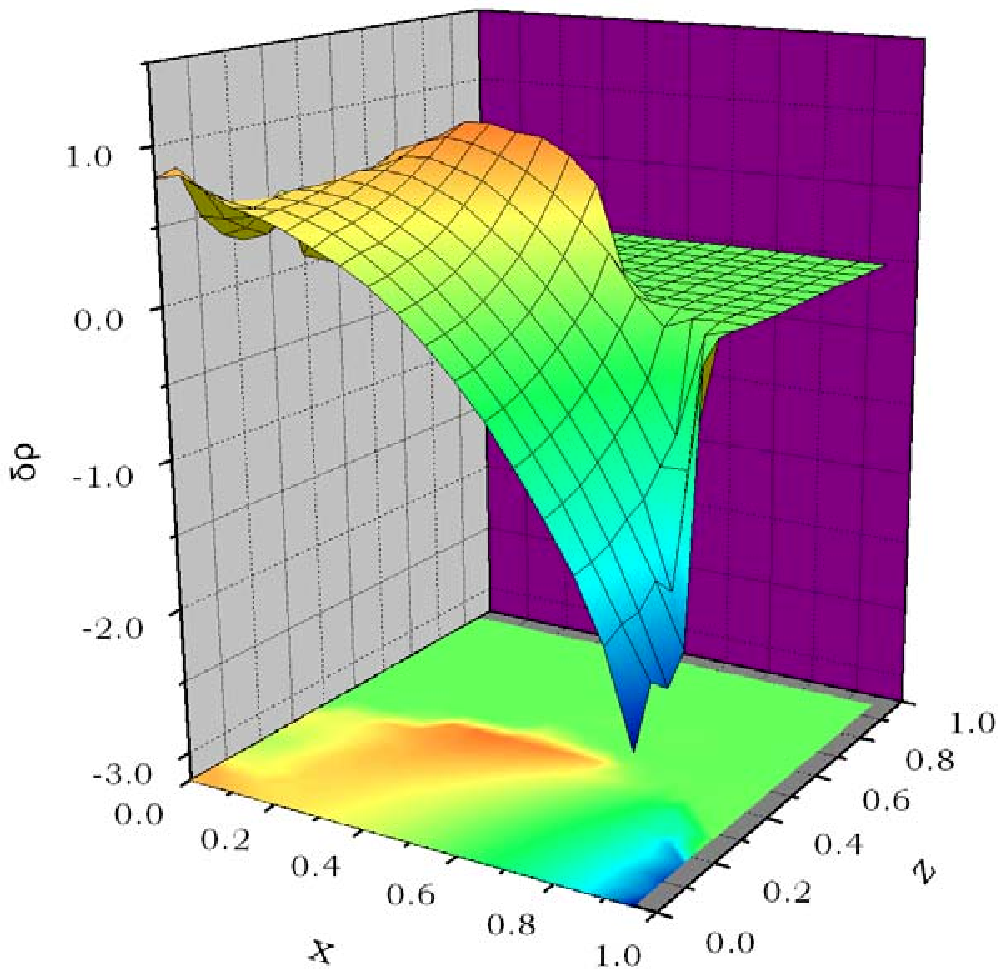}
\caption{Same as Figure \ref{fig:l2B0}, but for
  the most rapidly rotating model of sequence B (B12).}
\label{fig:l2B12}
\end{figure}

As discussed in Section \ref{sec:2} we obtain the two-dimensional
eigenfunction of a specific mode by computing Fourier transforms of
selected variables at every point inside the star. An example of such
an eigenfunction is shown in Figure \ref{fig:l2B0}, which displays the
eigenfunction corresponding to the density perturbations due to the
$l=2$ fundamental mode, for the nonrotating model B0. Since the model
is nonrotating, the eigenfunction shows the expected quadrupolar
structure. For the most rapidly rotating model B12, however, the
eigenfunction for the same mode appears severely altered (Figure
\ref{fig:l2B12}) with respect to the nonrotating limit. There are two
main effects caused by rapid rotation. Firstly, the $l=2$ fundamental
mode does not only couple to higher order spherical harmonics, but
also to an $l=0$ term. This causes a significant density variation at
the center of the star, whereas for the nonrotating model the center
of the star is not pulsating. Secondly, the centrifugal force weakens
the effective gravity in the equatorial region, causing the
eigenfunction to attain a large amplitude near the equatorial surface
of the star.  The effect of the centrifugal force becomes extreme when
the star rotates at the mass-shedding limit, as discussed in Section
\ref{sec:6}. The above effects are common in other modes, too, and
also in all four sequences considered.

\section{Splitting of the Fundamental mode}
\label{sec:5}

An interesting feature visible in Figures \ref{fig:freqsA} and
\ref{fig:freqsB} is the appearance of an additional fundamental mode,
which we denote as $F_{\rm II}$. This mode is degenerate in the
nonrotating limit, but becomes distinct from the $F$-mode for rotating
models. For sequence A, the $F_{\rm II}$-mode has a similar
dependence on rotation rate as the $F$-mode, with the difference in
frequency between the two modes remaining nearly constant with
increasing rotation rate. For sequence B, however, the frequency of
the $F_{\rm II}$-mode is decreasing with increasing rotation rate,
while the frequency of the $F$-mode is increasing.

\begin{figure}
\centerline{\psfig{file=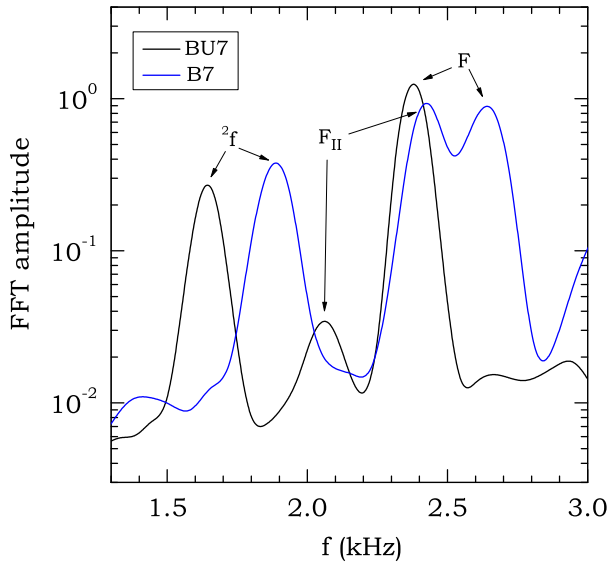,width=8cm}} 
\caption{Comparison of integrated Fourier amplitudes of the
  density evolution for a differentially rotating (B7) and a uniformly
  rotating (BU7) model. The peaks corresponding to the fundamental
  quasi-radial mode $F$, the additional fundamental mode $F_{\rm II}$ and
  the fundamental $l=2$ mode are shown.}
\label{fig:fft}
\end{figure}

\begin{figure}
\centerline{\psfig{file=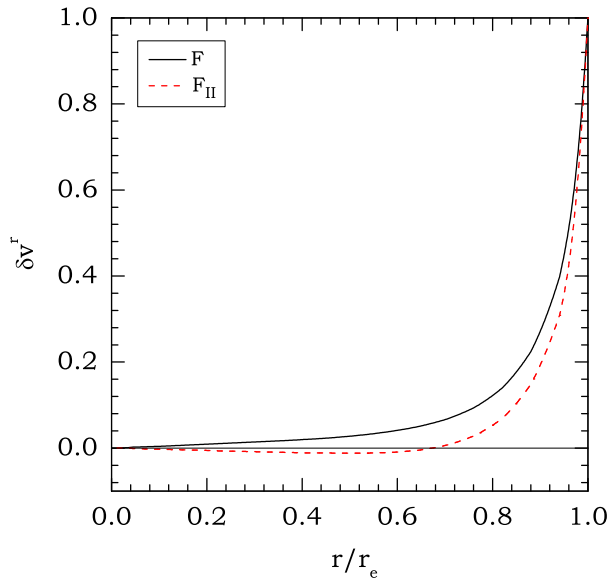,width=8cm}} 
\caption{Comparison of the 
  eigenfunctions, in the equatorial plane, corresponding to variations
  in the velocity component $v^r$ for the
  fundamental quasi-radial mode $F$ and for the additional fundamental
  mode $F_{\rm II}$ (for the uniformly rotating model BU7)}
\label{fig:eigFr}
\end{figure}

For both sequences A and B (which have a fixed strength of
differential rotation, $\hat A=1$) the $F_{\rm II}$-mode is excited at the
same level as the $F$-mode, by the perturbation shown in Eq.
(\ref{drho}) (the amplitude of the Fourier transform is similar for both
modes). Varying the degree of differential rotation (for the same
model and the same applied perturbation) we find that, while the amplitude of
the $F$-mode remains constant, the amplitude of the $F_{\rm II}$-mode
decreases as the star becomes less differentially rotating. In the
limit of uniform rotation, there still exists a peak in the Fourier
transform, corresponding to the $F_{\rm II}$ mode, although it has an
amplitude much smaller than the main $F$-mode (see Figure
\ref{fig:fft}, which compares the Fourier transforms for models B7 and
BU7). Extracting the eigenfunction of the radial velocity
perturbation, in the equatorial plane, for model BU7, shows that the
eigenfunction of the $F_{\rm II}$-mode has a node close to 70\% of the
equatorial radius, while the $F$-mode clearly has no node (see
Figure \ref{fig:eigFr}). Even though the $F_{\rm II}$-mode has a node
in the radial velocity, which is a property of a first overtone in
a nonrotating star, we still refer to it as a fundamental mode,
due to its degeneracy with the $F$-mode in a nonrotating star.

\begin{figure}
\includegraphics[width=8cm,clip]{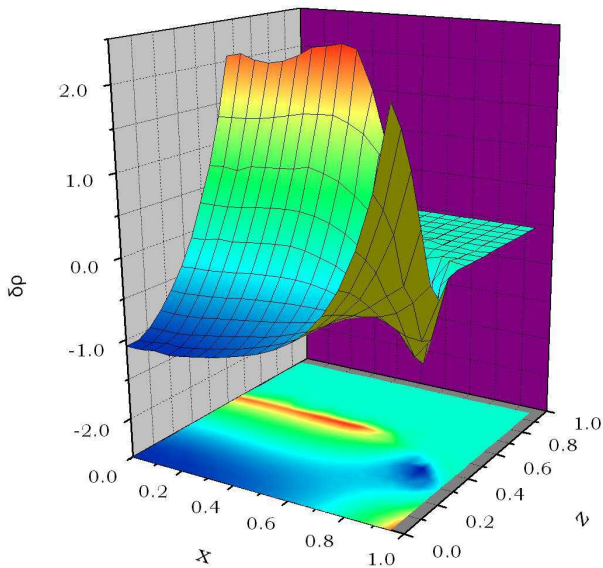}
\caption{Two-dimensional eigenfunction of the
  density perturbation, corresponding to the $l=0$ fundamental mode
  for the differentially rotating model B8.}
\label{fig:FB8}
\end{figure}

\begin{figure}
\includegraphics[width=8cm,clip]{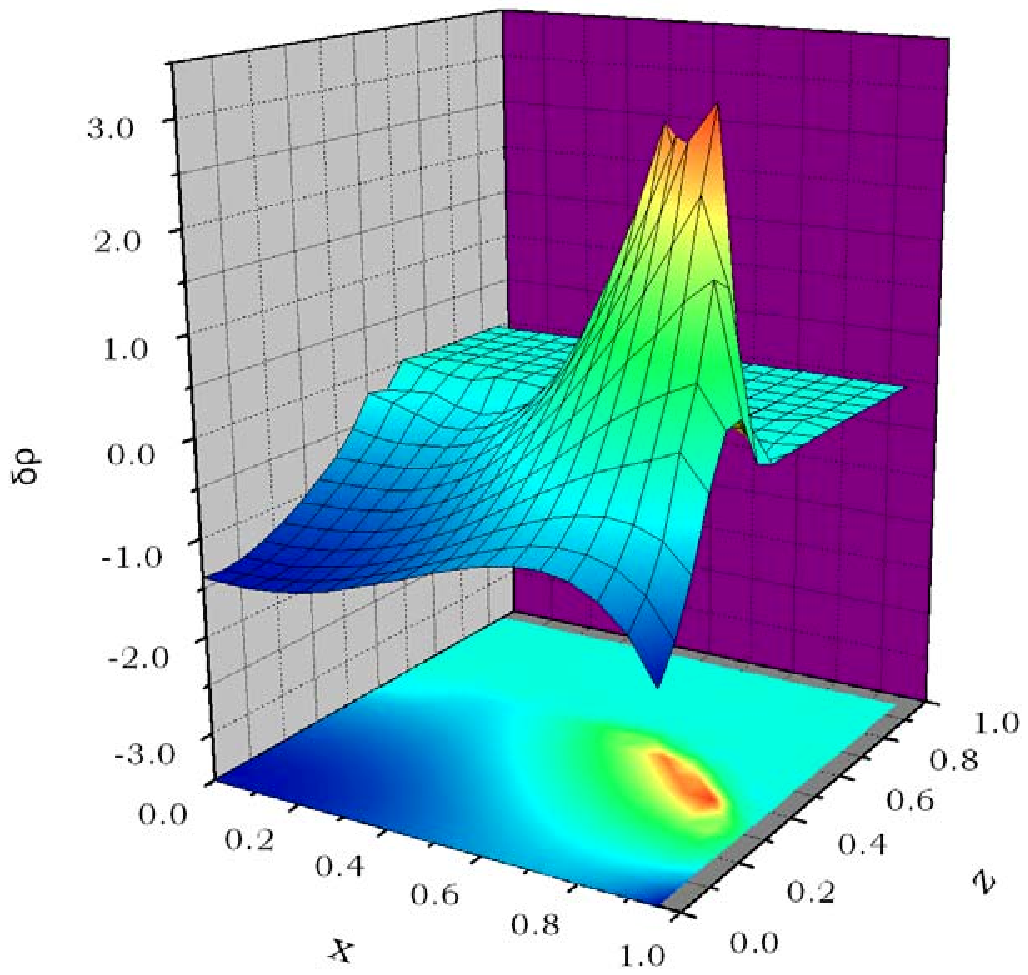}
\caption{Same as Figure \ref{fig:FB8}, but for
  the additional fundamental mode $F_{\rm II}$ (see discussion in Section
  \ref{sec:5}).}
\label{fig:FIIB8}
\end{figure}

The distinction between the $F$-mode and the $F_{\rm II}$-mode becomes
very clear in the extracted two-dimensional eigenfunctions. Figures
\ref{fig:FB8} and \ref{fig:FIIB8} show the eigenfunction of the
two modes, corresponding to the density perturbation, for the
differentially rotating model B8. The eigenfunction of the $F$-mode
is significantly modified by rotational couplings near the equatorial
surface, but otherwise it is similar to the expected eigenfunction
of the $F$-mode in a nonrotating star. In contrast, the eigenfunction
of the $F_{\rm II}$ mode is similar to an $F$-mode eigenfunction in
a nonrotating star only in the central region, but otherwise it
is very different in the polar and equatorial regions. The shape
of the eigenfunction in the equatorial plane (having no node in the
density perturbation) is reminiscent of density perturbations in
differentially rotating tori.

Is the $F_{\rm II}$-mode physical? Since our numerical code implements the
Cowling approximation we cannot answer this question at present. In
the Cowling approximation the energy and momentum conservation are
violated.  In previous simulations (Font et al. 2001) this violation
of the constraints has lead to the appearance of an unphysical
``fundamental'' dipole mode.  It is possible that the $F_{\rm II}$-mode
observed in the present simulations is also an artifact of the Cowling
approximation.  This can only be confirmed by new simulations in full
general relativity.

\section{Mass-shedding-induced damping of pulsations}
\label{sec:6}

Linear perturbations of rotating stars are assumed to have a
vanishingly small amplitude, so that the background equilibrium
star is unaffected by a linear oscillation mode. However, when
one considers a finite-amplitude oscillation, then nonlinear
effects can become important. Our nonlinear evolution code
allows us to investigate such effects.

At low rotation rates, a small oscillation amplitude of the order of
$10^{-2}$ does not lead to significant nonlinear effects. But, as the
star approaches the mass-shedding limit, the effective gravity near
the equatorial surface diminishes. Exactly at the mass-shedding limit
fluid elements are only marginally bound to the surface of the star. A
small radial pulsation then suffices to cause mass-shedding after each
oscillation period. At an oscillation frequency around 2.3 kHz (in the
Cowling approximation), matter that has been shed from the star does
not have the time to fall back before more matter is shed (this
statement applies to the specific example with an initial pulsation
amplitude of the order of $10^{-2}$, that is studied here -- more
generally, the fall-back time will depend on the initial amplitude).
In this way, a low-density toroidal envelope is created in the
equatorial region, which expands with every oscillation period.

\begin{figure}
  \centerline{\psfig{figure=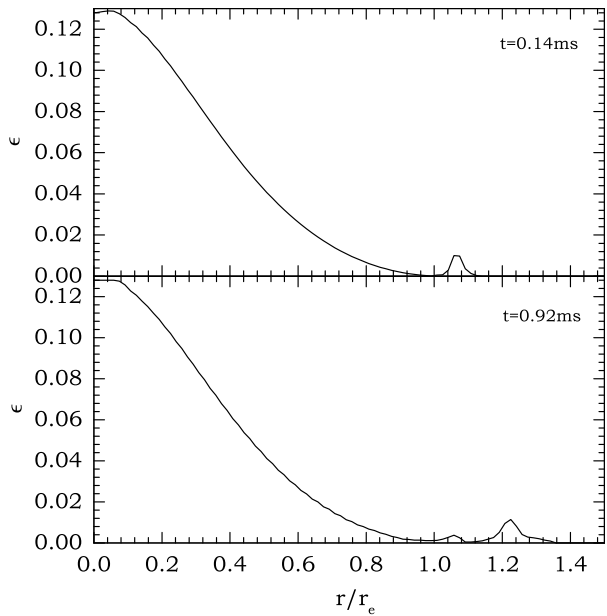,width=8cm}}
  \caption{Profiles of the internal energy $\epsilon$ showing the
    propagation of shocks, generated due to radial pulsations at the
    surface of the star. At $t=0.14$ms, a first shock is shown to
    propagate away from the star. At $t=0.92$ms, a second shock is
    propagating in the rarefaction region created by the first shock.
    A shock is produced after each period of the fundamental
    quasi-radial mode $F$. Pulsational energy is carried away by these
    shocks and a high-entropy envelope forms in the equatorial
    region surrounding the star.} \label{fig:shocks}
\end{figure}

Figure \ref{fig:shocks} shows the profile of the specific internal
energy $\epsilon$ at two different times, for the uniformly rotating
model BU9, which is at the mass-shedding limit. It is clearly seen that the
matter is shed in the form of high-entropy shock waves. A first shock
wave, shown at t=0.14ms, leaves the star as a result of the initial
perturbation applied to the equilibrium model. After one oscillation
period, a second shock wave leaves the star and both shock waves
are shown at t=0.92ms. In the same fashion, consecutive shock waves are
created after each oscillation period and the density in the toroidal
envelope gradually increases.

\begin{figure}
  \centerline{\psfig{file=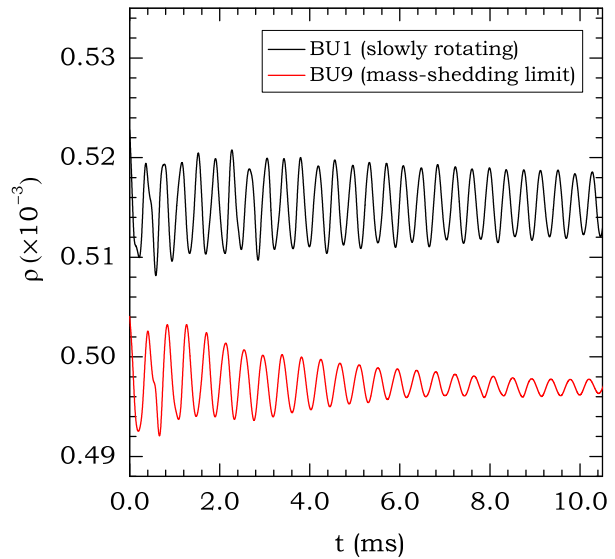,width=8cm}}
  \caption{Damping of nonlinear pulsations for a star rotating at the
    mass-shedding limit. For a slowly rotating star the pulsations are
    not damped (except by numerical viscosity). The time-evolution of
    the density is shown at half-radius in the star (for a better
    comparison, the lower curve was displaced by a fixed amount on the
    vertical axis). For an initial mode amplitude of $~1\%$, the
    nonlinear damping occurs on a dynamical timescale.}
\label{fig:damping}
\end{figure}

Since matter is shed from the star in the form of shocks, they carry
away kinetic energy, to the expense of the pulsational energy. In this
way, the pulsations of the star are gradually damped. Figure
\ref{fig:damping} shows a comparison between the time-evolution of the
central rest-mass density in the slowly rotating model BU1 and model
BU9. It is evident that the damping of the pulsations due to
mass-shedding is very strong.  The damping will also be effective in
stars which rotate at somewhat smaller rotation rates. The precise
damping rate will depend on several factors, such as the amplitude of
the pulsation, the rotation rate of the star, the EOS etc. The
assumption of the Cowling approximation which we have adopted in our
simulations will also affect the damping rate, as it causes an
imbalance between the different forces acting on a fluid element on
the equatorial surface of the star. Simulations which take into
account the time-evolution of the gravitational field during the
oscillations should yield more accurate damping rates.

Our finding of the mass-shedding induced damping of pulsations in
critically rotating stars can have severe consequences for unstable
pulsation modes, such as $f$-modes and $r$-modes, since, over a timescale
of several seconds, even a very small damping rate could suffice
to saturate their amplitude at values much less than order unity.

\section{Detectability of gravitational waves}
\label{sec:7}

Pulsating rotating neutron stars are gravitational-wave sources that
depend on several parameters (EOS, mass, angular
momentum, differential rotation law, initial amplitude, damping
mechanisms etc.).  All these parameters may have different effects on
the oscillation spectrum of the star and, therefore, the successful
extraction of the physical characteristics of the source from the
gravitational-wave signal will be difficult to achieve.  It is
important to isolate each effect on the gravitational waveform in
order to find general trends.  If independent information about the
source (a gravitational-wave burst, or optical, neutrino or gamma-ray
signals in the case of core collapse; a gravitational-wave chirp in
the case of a binary merger) can distinguish between isolated neutron
star formation and binary merger, then this could be used to constrain
the interesting range of several parameters.  Thus, a combined filter
that includes both a pre-formation characteristic signal and several
damped pulsation modes will enhance the total signal to noise ratio
(see Kokkotas, Apostolatos \& Andersson, 2001, for extraction of the
characteristic parameters of a damped monochromatic gravitational
waves, through matched filtering).

In the core collapse simulations by Dimmelmeier et al. (2001, 2002a,
2002b) the quasi-periodic gravitational waves emitted during core
collapse were found to have frequencies less than roughly 1.1 kHz and
our computed frequencies of $l=0$ and $l=2$ modes agree with this, as
discussed in Section \ref{fixed}. Such frequencies are still within
the range of current laser-interferometric detectors. However, the
above frequencies are typical only for certain EOSs.  The fundamental
$l=2$ $f$-mode frequency of the nonrotating model of sequence A is at
the lower end of the corresponding frequency range for 1.4$M_\odot$
models, constructed with a large sample of different realistic EOSs,
which range from $ \sim 1.35$ kHz (for extremely stiff EOSs) to $\sim
3.6$ kHz (for extremely soft EOSs), see e.g. Andersson \& Kokkotas
(1998). A similar range of frequencies, for different EOSs, exists for
the fundamental $l=0$ mode, for stars of 1.4$M_\odot$ mass. Based on
the empirical relations constructed in Section \ref{fixed}, we
estimate that, depending on the stiffness of the high-density EOS, 
the frequency range of quasi-periodic gravitational waves
during core collapse could be as low as $0.65$ to $1.35$ kHz for
extremely stiff EOSs and as high as $1.8$ to $3.6$ kHz for extremely
soft EOSs. The higher frequency range would be accessible for
gravitational-wave detection only with detectors such as the proposed
wide-band dual sphere (Cerdonio et al. 2001).


In the case of a neutron-star binary merger, the high-frequency
quasi-periodic oscillations excited in the hypermassive neutron star
could last for a large number of oscillation periods, since the
damping due to a low-density envelope should be much weaker than in the
core collapse case. Thus, the quasi-periodic signal could be enhanced
significantly by matched filtering.

\section{Discussion}
\label{sec:8}

Using an axisymmetric general relativistic hydrodynamics code we have
studied nonlinear pulsations of uniformly and differentially rotating
neutron stars. We have performed time-dependent numerical simulations
of a large sample of initial models which were slightly perturbed away
from hydrostatic equilibrium. The time-evolutions have been analyzed
using Fourier transforms at several points inside the stars, which
enables the extraction of the two-dimensional eigenfunction for each
mode. Our attention has been focused on two different sequences of
uniformly and differentially rotating stars, for which we have
obtained the pulsation frequencies for the two lowest-order
quasi-radial ($l=0$) and quadrupole ($l=2$) modes. We have found that
differentially rotating models can reach significantly lower pulsation
frequencies than uniformly rotating models of the same rest-mass. In
addition, our simulations show that the fundamental quasi-radial mode
is split into two different sequences. This new mode has been most
clearly observed in the fastest differentially rotating models. The
centrifugal forces and the degree of differential rotation have
significant effects on the mode-eigenfunctions. Most notably,
axisymmetric nonradial modes acquire a nonzero density variation at
the center of the star, due to coupling to lower-order terms in the
eigenfunction. However, both the splitting of the fundamental mode and the
nonzero central density variations of nonradial modes could be due
to the use of the Cowling approximation.

We have found that near the mass-shedding limit the pulsations are
damped due to shocks forming at the surface of the star, when matter
is shed in the equatorial region during each pulsation cycle. The
damping rate should depend on the amplitude of the pulsations and on
the rotation rate of the star. This new damping mechanism may set a
small saturation amplitude for modes that are unstable to the emission
of gravitational radiation. The nonlinear development of pulsations
when subject to this strong damping depends on the adopted EOS during
the time-evolution. If one restricts the perfect fluid to remain
isentropic, by assuming that the usual polytropic EOS holds throughout
the evolution, then this restriction does not allow for real shocks to
form and propagate away from the star, even though discontinuities
appear in the fluid variables in this case as well (actually, such
discontinuities are larger than the corresponding discontinuities in
the non-isentropic case, for the same mode-amplitude, due to kinetic
energy conservation).  Instead of propagating shocks, one obtains a
behaviour that is reminiscent of the ``wave-breaking'' of
large-amplitude $r$-modes in rapidly rotating stars, in nonlinear
simulations by Lindblom, Tohline \& Valisneri (2001)\footnote{We note
  that, in rapidly rotating stars, the $r$-mode velocity field has a
  significant radial component.}. In contrast, the correct approach is
to use a {\it nonisentropic} EOS during the evolution, such as the
ideal fluid EOS used in our simulations, which allows for physically
realistic shocks to form and propagate. We note that in the
simulations by Lindblom et al. (2001) the growth in the amplitude of
the unstable $r$-mode was accelerated by a large factor, leading to
the ``wave-breaking'' and the sudden destruction of the mode. We
conjecture that if one would use a nonisentropic EOS and would not
accelerate the natural $r$-mode growth time, then a balance between
the growth rate due to gravitational-radiation reaction and the
damping rate due to shock-dissipation caused by mass-shedding could be
obtained at some amplitude less than order unity. This discussion is
relevant, of course, only if in very rapidly rotating stars the
mass-shedding-induced damping limits the nonlinear amplitude of
unstable modes before other nonlinear saturation mechanisms can
dominate.  Examples of other nonlinear saturation mechanisms are
saturation due to the interaction of differential rotation with a
magnetic field (Rezzolla, Lamb \& Shapiro, 2000; Rezzolla et al.
2001a, 2001b), hydrodynamical instabilities of large-amplitude
nonlinear oscillations (Gressman et al. 2002) and nonlinear couplings
to other damped modes (Schenk et al. 2002; Morsink 2002; Arras et al.
2003). At present, it is still unclear what the relation, if any,
between the latter two mechanisms is (in the case of $r$-mode
oscillations). Which nonlinear saturation mechanism sets the maximum
amplitude of unstable modes could depend on the rotation rate and the
magnetic field strength of the star.

\section*{Acknowledgments}
It is a pleasure to thank Massimo Cerdonio, Harry Dimmelmeier, Kostas
Kokkotas, Jos\'e Pons, Hans Ritter, Henk Spruit and Arun Thampan
for helpful discussions.  We also thank the referee, Prof. L.
Rezzolla, for useful comments and Emanuele Berti for a careful reading
of the manuscript.  Financial support for this research has been
provided by the EU Network Programme (Research Training Network
Contract HPRN-CT-2000-00137).  T.A.A. acknowledges financial support
from the Special Accounts for Research Grants of the University of
Athens (grant 70/4/4056). J.A.F.  acknowledges financial support from
the Spanish Ministerio de Ciencia y Tecnolog\'{\i}a (grant AYA
2001-3490-C02-01). The computations were performed on the NEC SX-6 at
the Rechenzentrum Garching (Germany).


\begin{thebibliography}{0}

\bibitem{} Andersson, N., Kokkotas, K.~D., 1998, MNRAS, 299, 1059

\bibitem{} Arras, P., Flanagan, E.~E., Morsink, S.~M., Schenk, A.~K.,
Teukolsky, S.~A., Wasserman, I., 2003, Astrophys. J., 591, 1129 

\bibitem{} Baumgarte, T.~W., Shapiro, S.~L., Shibata, M., 2000,
  Astrophys. J., 528, L29

\bibitem{} Carroll, B.~W., Zweibel, E.~G., Hansen, C.~J., McDermott, P.~N., Savedoff, M.~P., Thomas, J.~H., Van Horn, H.~M., 1986, Astrophys. J., 305, 767

\bibitem{} Cerdonio, M., Conti, L., Lobo, J.~A., Ortolan, A.,
  Taffarello, L., Zendri, J.~P., 2001, Phys. Rev. Lett., 87, 031101

\bibitem{} Cook, J.~N., Shapiro, S.~L., Stephens, B.~C., 2003,
  astro-ph/0310304

\bibitem{} Dimmelmeier, H., Font, J.~A., M\"{u}ller, E., 2001,
  Astrophys. J., 560, L163

\bibitem{} Dimmelmeier, H., Font, J.~A., M\"{u}ller, E., 2002a,
 Astron. Astrophys., 388, 917

\bibitem{} Dimmelmeier, H., Font, J.~A., M\"{u}ller, E., 2002b,
 Astron. Astrophys., 393, 523

\bibitem{} Font, J.~A., Living Rev. Relativity, 2003, 6, 4

\bibitem{} Font, J.~A., Stergioulas, N., Kokkotas, K.~D., 2000,
  MNRAS, 313, 678

\bibitem{} Font, J.~A., Dimmelmeier, H., Gupta, A.,
Stergioulas, N., 2001, MNRAS, 325, 1463

\bibitem{} Font, J.~A., Goodale, T., Iyer, S., Miller, M.,
 Rezzolla, L., Seidel, E., Stergioulas, N., Suen, W.~M., Tobias, M., 2002,
 Phys.~Rev.~D, 65, 084024

\bibitem{} Friedman, J.~L., Lockitch, K.~H., 2001, in Proc. 9th Marcel
  Grossman Meeting, eds. Gurzadyan, V., Jantzen, R., Ruffini, R.,
  gr-qc/0102114

\bibitem{} Friedman, J.~L., Ipser, J.~R., Parker, L., 1986, Astrophys.
  J., 304, 115

\bibitem{} Fryer, C.~L., Holz, D~.E., Hughes, S.~A., Warren, M.~S., 2002,
astro-ph/0211609

\bibitem{} Gressman, P., Lin, L.-M., Suen, W.-M., Stergioulas, N.,
Friedman, J.~L., 2002, Phys. Rev. D., 66, 041303(R)

\bibitem{} Hartle, J.~B., Friedman, J.~L., Astrophys. J., 1975, 196, 653

\bibitem{} Heger, A., Woosley, S.~E., 2003, in Proc. Woods Hole, {\it
Gamma-Ray Bursts and Afterglow Astronomy} (AIP), in press

\bibitem{} Heger, A., Woosley, S.~E., Langer, N., Spruit, H.~C., 2003, in
Proc. IAU Symp. 215, {\it Stellar Rotation}, eds. Maeder, A., Eenens, P.,
astro-ph/0301374

\bibitem{} Hegyi, D.~J, 1977, Astrophys. J., 217, 244

\bibitem{} Ivanova, N., Podsiadlowski, Ph., 2003, in {\it From
    Twilight to Highlight: the Physics of Supernovae}, eds.
  Hilledrandt, W., Leibundgut, B. (Springer, Berlin) p. 19

\bibitem{} Jones, P~.B., 2001, Phys. Rev. Lett., 86, 1384

\bibitem{} Kokkotas, K.~D., Andersson, N., 2001, in Proc.
SIGRAV XIV, Genoa 2000, Springer-Verlag, gr-qc/0109054

\bibitem{} Kokkotas, K.~D., Apostolatos, T.~A., Andersson, N., 2001
 MNRAS, 320, 307

\bibitem{} Kokkotas, K.~D., Schmidt, B., 1999, Living Rev. Relativity, 2, 2

\bibitem{} Komatsu, H., Eriguchi, Y., Hachisu, I., 1989a, MNRAS,
 237, 355

\bibitem{} Komatsu, H., Eriguchi, Y., Hachisu, I., 1989b, MNRAS,
 239, 153

\bibitem{} Kotake, K., Yamada, S., Sato, K., 2003, Phys. Rev. D, 68, 044023

\bibitem{} Langer, N., Yoon, S.-C., Petrovic, J., Heger, A., 2003,
  astro-ph/0302232

\bibitem{} Lai, D., Shapiro, S.~L., 1995, Astrophys. J., 442, 259

\bibitem{} Lindblom, L., Owen, B.~J., 2002, Phys. Rev. D, 65, 063006

\bibitem{} Lindblom, L., Tohline, J.~E., Vallisneri, M., 2001, 
Phys. Rev. Lett., 86, 1152

\bibitem{} Liu, Y.~T., Lindblom, L., 2000, MNRAS,
 324, 1063

\bibitem{} Liu, Y.~T., Shapiro, S.~L., 2003, gr-qc/0312038

\bibitem{} McDermott, P.~N., Savedoff, M.~P., Van Horn, H.~M., Zweibel, E.~G.,
Hansen, C.~J., 1984, Astrophys. J., 281, 746

\bibitem{} Middleditch, J., 2003, astro-ph/0311484

\bibitem{} Morsink, S.~M., 2002, Astrophysical J., 571, 435

\bibitem{} Morsink, S.~M., Stergioulas, N., Blattnig, S.~R., 1999
Astrophys. J., 510, 854

\bibitem{} M\"onchmeyer, R., Sch\"afer, G., M\"uller, E., Kates, R.~E.,
1991, Astron. Astrophys., 246, 417

\bibitem{} New, K., 2003, Living Rev. Relativity, 6, 2

\bibitem{} Nice, D.~J., Splaver, E.~M., Stairs, I.~H., 2003,
  astro-ph/0311296

\bibitem{} Nozawa, T., Stergioulas, N., Gourgoulhon, E., Eriguchi, Y.,
1998, Astron. Astrophys. Suppl., 132, 431

\bibitem{} Ott, C.~D, Burrows, A., Livne, E, Walder, R., 2003,
astro-ph/0307472

\bibitem{} Podsiadlowski, Ph., Langer, N., Poelarends, A.~J.~T.,
  Rappaport, S., Heger, A., Pfalh, E, 2003, astro-ph/0309588

\bibitem{} Pons, J.~A., 2003, personal communication

\bibitem{} Pfahl, E., Rappaport, S., Podsiadlowski, Ph., Spruit, H., 2002,
Astrophys. J., 574, 364

\bibitem{} Rezzolla, L., Lamb, F.K., Shapiro, S.L., 2000
Astrophys. J., 531, L139

\bibitem{} Rezzolla, L., Lamb, F.K., Markovi{\'c}, D., Shapiro, S.L., 2001a
Phys. Rev. D, 64, 104013

\bibitem{} Rezzolla, L., Lamb, F.K., Markovi{\'c}, D., Shapiro, S.L., 2001b
Phys. Rev. D, 64, 104014

\bibitem{} Schenk, A.~K., Arra, P., Flanagan, \'E.~\'E., Teukolsky, S.~A.,
Wasserman, I., 2002, Phys. Rev. D., 65, 024001

\bibitem{} Shapiro, S.~L., 2000, Astrophys. J., 544, 397

\bibitem{} Shibata, M., 2003, Phys. Rev. D, 67, 024033

\bibitem{} Shibata, M., Taniguchi, K., Ury\={u}, K., 2003, Phys.~Rev.~D,
 68, 084020

\bibitem{} Shibata, M., Ury\={u}, K., 2000, Phys.~Rev.~D,
 61, 064001

\bibitem{} Shibata, M., Ury\={u}, K., 2002, Prog. Theor.
  Phys., 107, 265

\bibitem{} Sperhake, U., 2002, PhD Thesis, gr-qc/0201086

\bibitem{} Sperhake, U., Papadopoulos, P., Andersson, N., 2001, gr-qc/0110487

\bibitem{} Spruit, H.~C., 2002, Astron. Astrophys., 381, 923

\bibitem{} Stergioulas, N., 2003, Living Rev. Relativity, 6, 3

\bibitem{} Stergioulas, N., Friedman, J.~L., Astrophys. J., 1995,
444, 306

\bibitem{} Stergioulas, N., Friedman, J.~L., Astrophys. J., 1998,
492, 301

\bibitem{} Stergioulas, N., Font, J.~A., 2001, Phys. Rev. Lett.,
86, 1148

\bibitem{} van Dalen, E.~N.~E., Dieperink, A.~E.~L., 2003, nucl-th/0311103

\bibitem{} Villain, L., Pons, J. A., Cerd\'a-Dur\'an, P.,
  Gourgoulhon, E., 2003, astro-ph/0310875

\bibitem{} Watts, A.~L., Andersson, N., 2002, MNRAS, 333, 943

\bibitem{} Yoshida, S., Eriguchi, Y., 1999, Astrophys. J., 515, 414

\bibitem{} Yoshida, S., Eriguchi, Y., 2001, MNRAS, 322, 389

\bibitem{} Yoshida, S., Kojima, Y., 1997, MNRAS, 289, 117

\bibitem{} Yoshida, S., Rezzolla, L., Karino, S., Eriguchi, Y., 2002
Astrophys. J., 568, L41

\bibitem{} Yuan, Y., Heyl, J.~S., 2003, astro-ph/0305083

\bibitem{} Zwerger, T., M\"uller, E., 1997, 320, 209

\end{thebibliography}
\end{document}